\documentclass[a4paper,11pt]{article}

\usepackage{jcappub}
\usepackage[T1]{fontenc}
\usepackage{lineno}

\usepackage{float}

\title{\boldmath A Log-Normal Timing Likelihood for Direction Reconstruction of Ultra-High-Energy Gamma-Ray Air Showers}

\author[a,b]{Jun-Yu Shen,}
\author[a,b,1]{Yi Zhang,\note{Corresponding author.}}
\author[a,b]{Jia-Yin He,}
\author[a,b]{Xing-Yu Li,}
\author[a,b]{Shi-Ping Zhao,}
\author[a,b,2]{and Qiang Yuan\note{Corresponding author.}}

\affiliation[a]{Key Laboratory of Dark Matter and Space Astronomy, Purple Mountain Observatory, Chinese Academy of Sciences, Nanjing 210023, People's Republic of China}
\affiliation[b]{School of Astronomy and Space Science, University of Science and Technology of China, Hefei, Anhui 230026, People's Republic of China}

\emailAdd{zhangyi@pmo.ac.cn}
\emailAdd{yuanq@pmo.ac.cn}

\abstract{
	Ultra-high-energy $\gamma$-ray observations provide a direct probe of Galactic PeVatrons. Extending measurements to multi-PeV energies requires characterizing extensive air showers over large lateral distances, where the arrival-time distribution of secondary particles becomes broad and strongly asymmetric. Conventional direction reconstruction methods approximate the shower front as a deterministic surface and minimize a $\chi^2$ function of timing residuals. Because this implicitly assumes symmetric Gaussian errors, far-axis timing information is poorly utilized. \textsc{CORSIKA} simulations show that the arrival-time distribution of $\gamma$-induced secondary particles in the shower plane at a given lateral distance is positively skewed and closely follows a log-normal function. Using this property, we develop a likelihood-based direction reconstruction method that directly models the empirical arrival-time distribution and its radial dependence while accounting for multiple particles triggering the same detector unit. Applied to simulated $\gamma$-induced air showers, the log-normal likelihood consistently improves angular resolution compared with conical and parabolic wavefront fits, with the largest relative gains for high-energy events spanning wide lateral extents. We also show that independent readout of individual detector units yields better angular resolution than merging their signals into a single station-level measurement.
}

\begin{document}
	
	\maketitle
	\flushbottom
	
	\section{Introduction}
	
	Identifying and characterizing Galactic PeVatrons and Super-PeVatrons, astrophysical sources capable of accelerating cosmic rays (CRs) to PeV and multi-PeV energies, are central goals of modern high-energy astrophysics \citep{Gabici2019, Anguner2023, CG2023}. Because ultra-high-energy (UHE) $\gamma$ rays trace the hadronic interactions of these CRs and propagate undeflected through Galactic magnetic fields, they provide a direct probe of the accelerators \citep{Cao2023}. Although recent observations have confirmed the existence of Galactic PeVatrons, higher sensitivity in the multi-PeV band is needed to expand the source sample, map extended emission, and constrain the underlying acceleration mechanisms \citep{Aharonian2021, Luque2023}.
	
	Observations of UHE $\gamma$ rays rely primarily on large ground-based extensive air-shower (EAS) arrays. A primary UHE $\gamma$ ray entering the atmosphere initiates an electromagnetic cascade of secondary particles (SPs), a fraction of which reach ground level and are sampled by surface detectors to reconstruct the primary arrival direction \citep{Cao2023}. Current arrays, notably LHAASO-KM2A, have detected numerous UHE $\gamma$-ray sources and extended measurements up to $\sim 3.7~\mathrm{PeV}$ \citep{Cao2024,LHAASO2024Bubble,LHAASO2026CygnusX3,LHAASO2026Peanut,LHAASO2026PSRJ1849}. Probing the maximum energies attainable by Galactic accelerators requires pushing observations even higher, where the steeply falling $\gamma$-ray flux demands a much larger effective area. Future arrays therefore favor sparse layouts with wide inter-station spacing to keep construction costs manageable. Such sparse configurations reduce the number of triggered detectors per event and force direction reconstruction to rely on measurements at large lateral distances, where intrinsic shower-front fluctuations become pronounced \citep{Linsley1985,Bonifazi2009} and limit the achievable angular resolution.
	
	Good angular resolution is vital for EAS-based $\gamma$-ray astronomy, both to suppress the dominant hadronic CR background in point-source searches \citep{HAWC2025} and to resolve the morphology of extended sources \citep{HESS2018,Cao2021}. The achievable angular resolution depends directly on how accurately the geometry and timing structure of the shower front are modeled. Previous studies have shown that the shower front has a finite temporal thickness that grows with lateral distance $r_{\perp}$ as path-length differences among SPs increase \citep{Inoue1985,Linsley1986,Agnetta1997,Huang2019}. In addition, the shower front is curved, because SPs at larger $r_{\perp}$ travel longer geometrical paths and arrive systematically later than expected for a plane front \citep{Bassi1953,Zeng2025}. Measured arrival times thus exhibit both a distance-dependent delay from curvature and an intrinsic spread from the finite shower-front thickness.
	
	To account for these timing properties, ground-based experiments have developed various parameterized shower-front models. Examples include the modified Linsley function adopted by the Telescope Array \citep{Teshima1986, Ivanov2021}, plane-front fits with curvature corrections used by the Pierre Auger Observatory, HAWC, and Milagro \citep{Aab2014,Milagro,Atkins2004,HAWCPass5}, and conical or quadratic wavefront models employed by LHAASO and the Tibet AS$\gamma$ array \citep{Aharonian2021,Zeng2025,TibetConeFit}. KASCADE-Grande instead uses a two-step scheme that refines an initial plane-front direction with a simulation-based shower-front description \citep{KASCADEGrande,Antoni2004KASCADE}. Most of these methods treat the shower front as a deterministic geometric surface and estimate the arrival direction by minimizing a $\chi^2$ function of timing residuals, implicitly assuming symmetric Gaussian errors. This approximation holds reasonably well when a detector unit registers many SPs, as the earliest arrival time is nearly Gaussian distributed. At large lateral distances, however, the expected number of SPs per unit drops below unity, and the arrival-time distribution becomes strongly non-Gaussian. To suppress these asymmetric fluctuations, conventional fits rely on strong spatial and temporal weighting, which limits the use of far-axis timing information. Detailed simulations show that the arrival-time distribution of $\gamma$-induced SPs in the shower plane at a given $r_{\perp}$ is positively skewed and closely follows a log-normal function \citep{Lafebre2009,Bahmanabadi2018,Hofmann2020}. Constructing a timing likelihood directly from this distribution makes more efficient use of the available measurements, which is essential for sparse arrays that rely on limited far-axis sampling.
	
	Based on these timing properties, we develop a log-normal timing likelihood for direction reconstruction in ground-based EAS arrays. The method models the empirical arrival-time distribution and its radial dependence directly, and accounts for multiple particles triggering a single detector unit. We evaluate its performance with Monte Carlo simulations of UHE $\gamma$-induced showers in a sparse surface array and compare it with conventional conical and parabolic wavefront fits. Section 2 describes the simulation framework, Section 3 details the reconstruction models, Section 4 presents the angular resolution results and comparisons, and Section 5 summarizes our conclusions and discusses future extensions.
	
	\section{Monte Carlo Simulations}
	
	We performed Monte Carlo simulations to evaluate the direction reconstruction performance of a sparse EAS array composed of electromagnetic particle detectors (EDs) for UHE $\gamma$-ray events. The simulation chain consisted of two stages: air-shower generation with \textsc{CORSIKA} and detector-response simulation with \textsc{GEANT4}.
	
	\subsection{Air Shower Simulation}
	
	Air showers initiated by primary $\gamma$ rays were simulated with \textsc{CORSIKA} 7.7550 \citep{corsika}, using URQMD 1.3cr \citep{Heck1998} for low-energy hadronic interactions and SIBYLL 2.3d \citep{Riehn2020} for high-energy interactions. Only primary $\gamma$ rays were simulated in this work, with a total of 100,000 showers generated at an observation altitude of $3~\mathrm{km}$ above sea level. Primary energies were sampled from $100~\mathrm{TeV}$ to $30~\mathrm{PeV}$ following a differential power-law spectrum, $\mathrm{d}N/\mathrm{d}E \propto E^{-1.5}$, to ensure sufficient statistics at the highest energies, and the events were subsequently reweighted to a differential power-law spectrum of $\mathrm{d}N/\mathrm{d}E \propto E^{-2.7}$. Zenith angles were sampled uniformly in $\cos\theta$ from $0^\circ$ to $50^\circ$, and azimuth angles were sampled uniformly from $0^\circ$ to $360^\circ$. For each shower, the positions, arrival times, and energies of all SPs reaching the observation level were recorded as input for the detector simulation.
	
	\subsection{Detector Array and Response Simulation}
	
\begin{figure}
		\centering
		\includegraphics[width=0.6\linewidth]{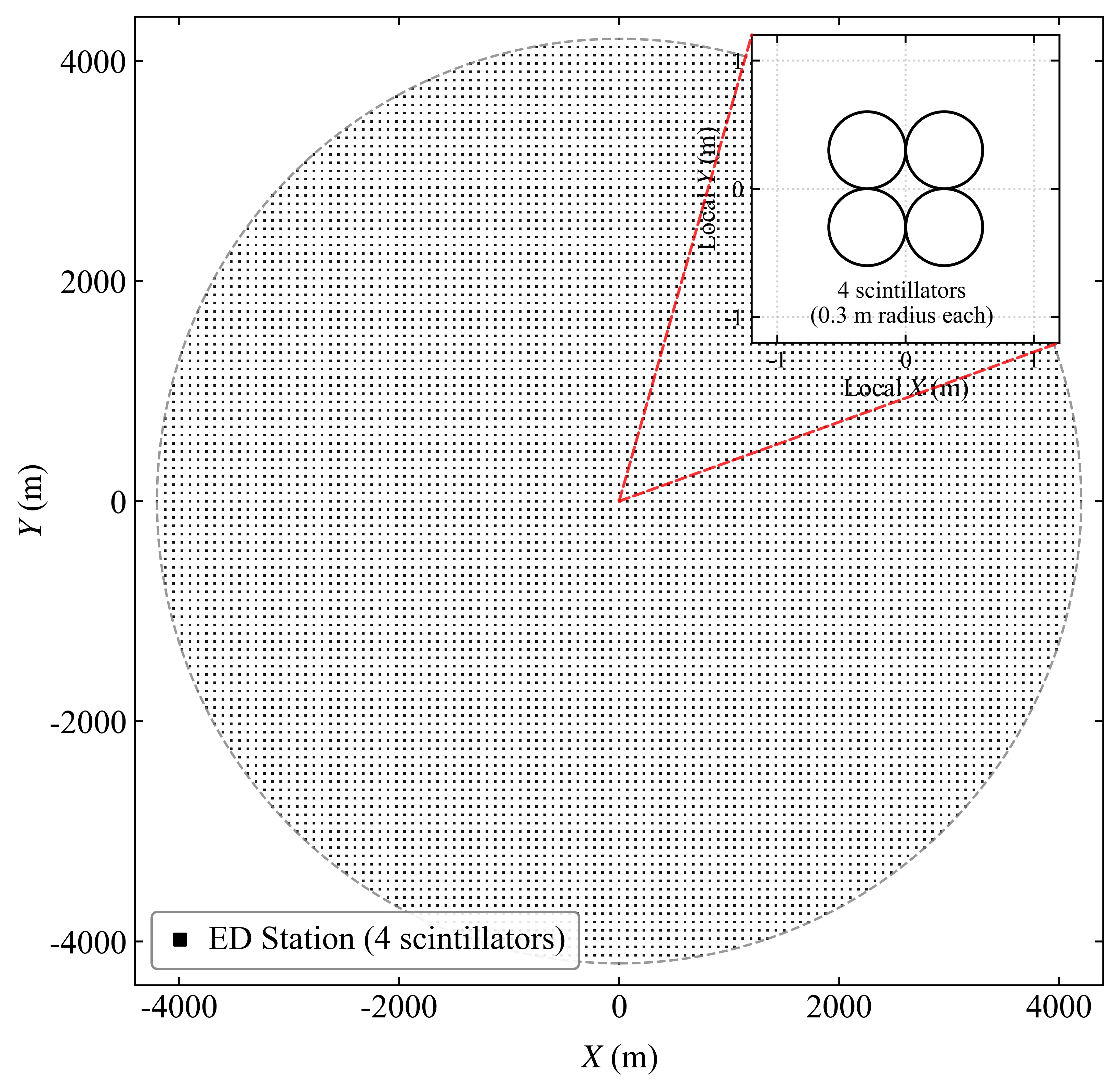}
		\caption{Layout of the simulated detector array. The main panel shows the full array of detector stations (black squares) arranged on a square grid with a spacing of $75\,\mathrm{m}$, confined within a circular area of radius $4.2\,\mathrm{km}$ (indicated by the grey dashed circle). Each station consists of four closely packed scintillator units, as illustrated in the inset: circular scintillators of $0.3\,\mathrm{m}$ radius in a $2\times2$ configuration.}
		\label{fig:array}
	\end{figure}

	The simulated sparse ED array comprises approximately 10,000 detector stations arranged on a square grid with a spacing of $75\,\mathrm{m}$, covering a circular area of radius $4.2\,\mathrm{km}$ centered at the origin. Each station consists of four cylindrical liquid-scintillator units with a radius of $0.3\,\mathrm{m}$ and a thickness of $6\,\mathrm{cm}$, arranged in a $2\times2$ configuration (see Figure~\ref{fig:array}). Shower cores were sampled uniformly in area within a central circular region of radius $1\,\mathrm{km}$, keeping all events well inside the array to avoid boundary effects.
	
	We simulated the detector response using \textsc{GEANT4} (version 11.2.0) \citep{geant4} by propagating the \textsc{CORSIKA} SPs through the scintillator geometry. For each unit, the earliest trigger time and the total deposited energy $E_{\rm dep}$ were recorded. To account for instrumental timing limits, including detector time resolution and station synchronization, we applied a Gaussian smearing of $3~\mathrm{ns}$ to the simulated trigger time of each unit.
	
	From our simulations, the most probable energy deposited by a minimum-ionizing particle (MIP) traversing a $6\,\mathrm{cm}$ thick liquid-scintillator unit is $E_{\rm MIP} = 10~\mathrm{MeV}$. Only units with $E_{\rm dep} > 0.5\,E_{\rm MIP}$ were retained for reconstruction, suppressing sub-threshold deposits from low-energy SPs while preserving the MIP signal. Assuming approximate proportionality between $E_{\rm dep}$ and the number of MIP equivalents, we define an effective multiplicity $N$ as
	\begin{equation}
		\label{eq:effective_multiplicity}
		N = \frac{E_{\rm dep}}{E_{\rm MIP}},
	\end{equation}
	which serves as a continuous estimate of the number of SPs contributing to the signal in each unit throughout the direction reconstruction.
	
	\subsection{Trigger Condition and Event Selection}
	
	A detector station was considered triggered if at least one of its four scintillator units satisfied $E_{\rm dep} > 0.5\,E_{\rm MIP}$, and the total number of triggered stations in an event is denoted by $N_{\rm TrigE}$. To ensure sufficient timing measurements for direction reconstruction, only events with $N_{\rm TrigE} > 9$ were retained. After this selection, approximately 48,000 events remained for analysis.
	
	\section{Direction Reconstruction Models}
	
	This section describes the three direction reconstruction models compared in this work: conical and parabolic wavefront fits, and a log-normal timing likelihood model. All three models operate on the same set of triggered scintillator units.
	
	\subsection{General Framework}
	
	All three models determine the shower arrival direction by minimizing an objective function $\mathcal{F}$ constructed from the timing residuals of the triggered units. For a trial shower direction unit vector $\hat{n}$, the timing residual of the $i$-th unit is written as
	\begin{equation}\label{eq:general_residual}
		\delta t_i = t_i - t_{\text{pred},i}
		= t_i - t_{0} - \frac{1}{c}
		\left(\vec{r}_i \cdot \hat{n} + \eta\, f(r_{\perp,i})\right),
	\end{equation}
	where $t_{\text{pred},i}$, $t_i$, $t_{0}$, and $c$ are the model-predicted arrival time, the measured arrival time, a common reference time, and the speed of light, respectively. Here, $\vec{r}_i$ denotes the position of the scintillator unit relative to the shower core in the ground frame, $r_{\perp,i}$ is the corresponding lateral distance in the shower plane, $\eta$ is a model-dependent shape parameter, and $f(r_{\perp,i})$ specifies the radial dependence of the arrival-time delay. Using these residuals, we define a per-unit contribution $\rho_i(\delta t_i)$ and obtain the reconstructed direction by minimizing
	\begin{equation}\label{eq:general_objective}
		\mathcal{F} = \sum_i \rho_i(\delta t_i).
	\end{equation}
	The three models differ in the form of the radial term $f(r_{\perp,i})$, the per-unit contribution $\rho_i(\delta t_i)$, and the treatment of $t_0$.
	
	\subsection{Conical Wavefront Model}
	
	The conical wavefront model approximates the shower front as a conical surface whose arrival-time delay increases linearly with $r_{\perp}$ from the shower axis. Setting $f(r_{\perp,i}) = r_{\perp,i}$ and $\eta = \alpha$, the predicted arrival time at the $i$-th triggered unit is
	\begin{equation}\label{eq2}
		t_{\text{pred},i} = t_{0} + \frac{1}{c}\left(\vec{r}_i \cdot \hat{n}
		+ \alpha\, r_{\perp,i}\right),
	\end{equation}
	where $\alpha$ is the conical curvature parameter. In this model, the direction vector $\hat{n}$, the reference time $t_0$, and the curvature parameter $\alpha$ are fitted simultaneously.
	
	The shower direction is determined by minimizing
	\begin{equation}\label{eq:chi2_conical}
		\mathcal{F} = \sum_i \rho_i(\delta t_i)
		= \sum_i w^2(r_{\perp,i})\,\zeta^2(\delta t_i)\,\delta t_i^2,
	\end{equation}
	which corresponds to the $\chi^2$ function used in Aharonian et al. \citep{Aharonian2021}. The spatial weight is
	\begin{equation}\label{eq:spatial_weight}
		w(r_{\perp,i}) = \frac{1}{r_{\perp,i}},
	\end{equation}
	where a small constant term is added to the denominator in practice to prevent divergence as $r_{\perp,i} \to 0$. The asymmetric temporal weight is
	\begin{equation}\label{eq:time_weight}
		\zeta(\delta t_i) =
		\begin{cases}
			1, & -50~\mathrm{ns} < \delta t_i < 150~\mathrm{ns},\\[4pt]
			\exp\!\left[-\frac{1}{2}\left(\frac{\delta t_i}{10~\mathrm{ns}}\right)^2\right],
			& \delta t_i \ge 150~\mathrm{ns} \text{ or } \delta t_i \le -50~\mathrm{ns},
		\end{cases}
	\end{equation}
	which strongly suppresses timing residuals outside the nominal window.
	
	\subsection{Parabolic Wavefront Model}
	
\begin{figure}
		\centering
		\includegraphics[width=0.65\linewidth]{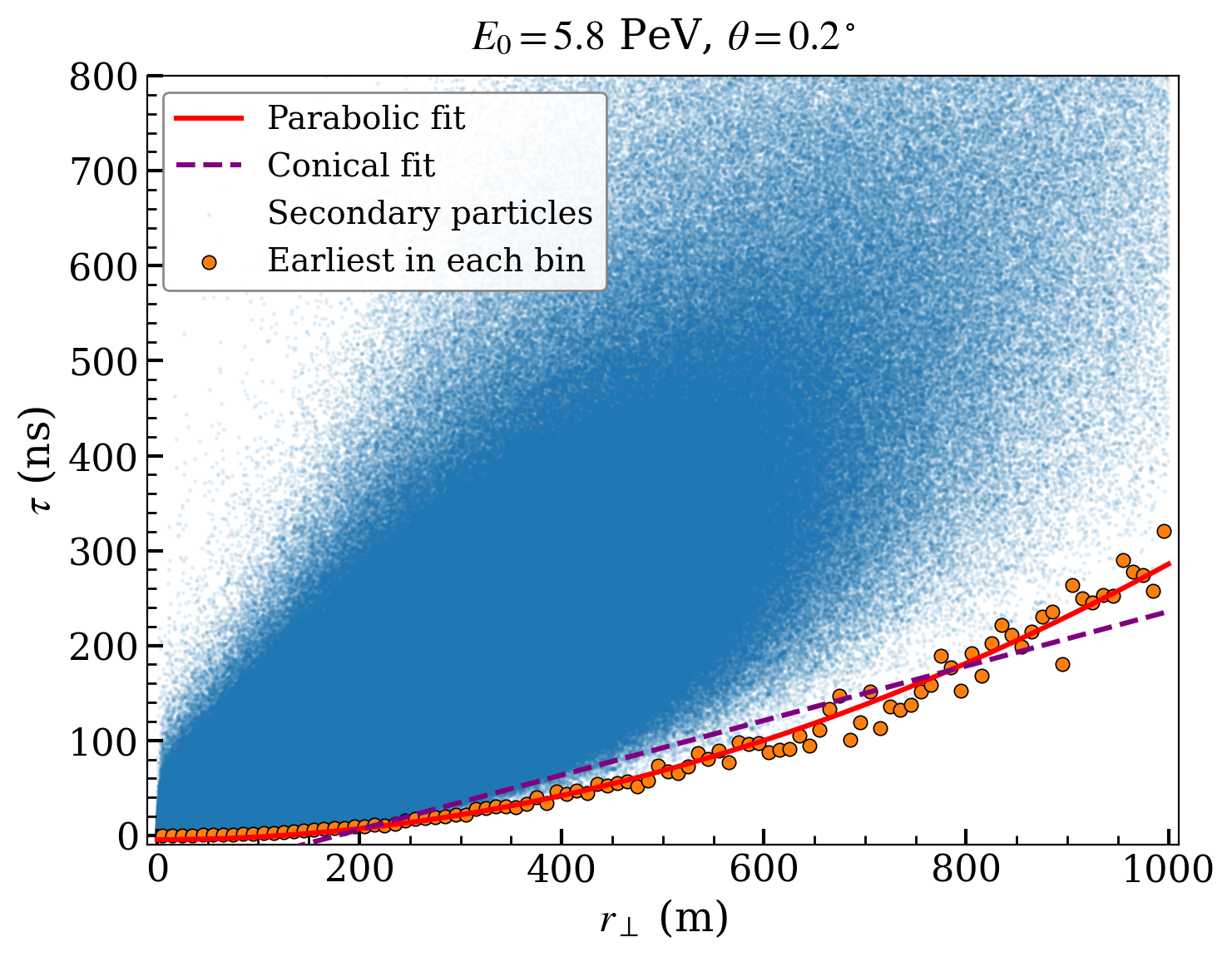}
		\caption{Arrival-time residuals $\tau$ of electromagnetic SPs in the shower plane plotted against the lateral distance $r_\perp$, for a representative $\gamma$-induced air shower with primary energy $E_0 \simeq 5.8~\mathrm{PeV}$ and zenith angle $\theta \simeq 0.2^\circ$. Each blue point represents one SP, with $\tau = t_{\rm sp} - t_{\rm base}$, where $t_{\rm sp}$ is the projected arrival time in the shower plane and $t_{\rm base}$ is the earliest $t_{\rm sp}$ over this plane. Orange circles mark the earliest-arriving particle in each $10~\mathrm{m}$ wide radial bin, tracing the leading edge of the shower front. The purple dashed curve shows a linear fit ($\tau \propto r_\perp$), while the red solid curve shows a parabolic fit ($\tau \propto r_\perp^2$), which describes the leading edge well at large $r_\perp$.}
		\label{fig:shape}
	\end{figure}

	To assess whether a higher-order curvature correction is needed beyond the conical approximation, we examine the shower-front timing structure directly from the \textsc{CORSIKA} simulations. Figure~\ref{fig:shape} shows the arrival-time residuals $\tau$ of SPs in the shower plane as a function of $r_\perp$ for a representative $\gamma$-induced air shower with primary energy $E_0 \simeq 5.8~\mathrm{PeV}$ and zenith angle $\theta \simeq 0.2^\circ$. Here, $\tau$ is defined as $t_{\rm sp} - t_{\rm base}$, where $t_{\rm sp}$ is the projected arrival time of each SP in the shower plane and $t_{\rm base}$ is the earliest $t_{\rm sp}$ over the plane. Recording the earliest arrival time in each $10~\mathrm{m}$ wide radial bin (orange circles) traces the leading edge of the shower front. This profile departs clearly from a linear relation and is well described by a quadratic dependence, $\tau \propto r_\perp^2$, indicating that a purely linear parameterization does not adequately capture the leading edge at large $r_\perp$.
	
	Motivated by this behavior, we adopt a parabolic wavefront model with $f(r_{\perp,i}) = r_{\perp,i}^2$ and $\eta = \beta$, giving
	\begin{equation}\label{eq4}
		t_{\text{pred},i} = t_{0} + \frac{1}{c}\bigl(\vec{r}_i \cdot \hat{n}
		+ \beta\, r_{\perp,i}^{2}\bigr),
	\end{equation}
	where $\beta$ is the parabolic curvature parameter. As in the conical model, $\hat{n}$, $t_0$, and $\beta$ are treated as free parameters and determined by minimizing the same objective function $\mathcal{F}$ with the same spatial and temporal weights.
	
	\subsection{Log-Normal Timing Likelihood}
	
	\subsubsection{Motivation}
	
\begin{figure}[htbp]
		\centering
		\includegraphics[width=0.65\textwidth]{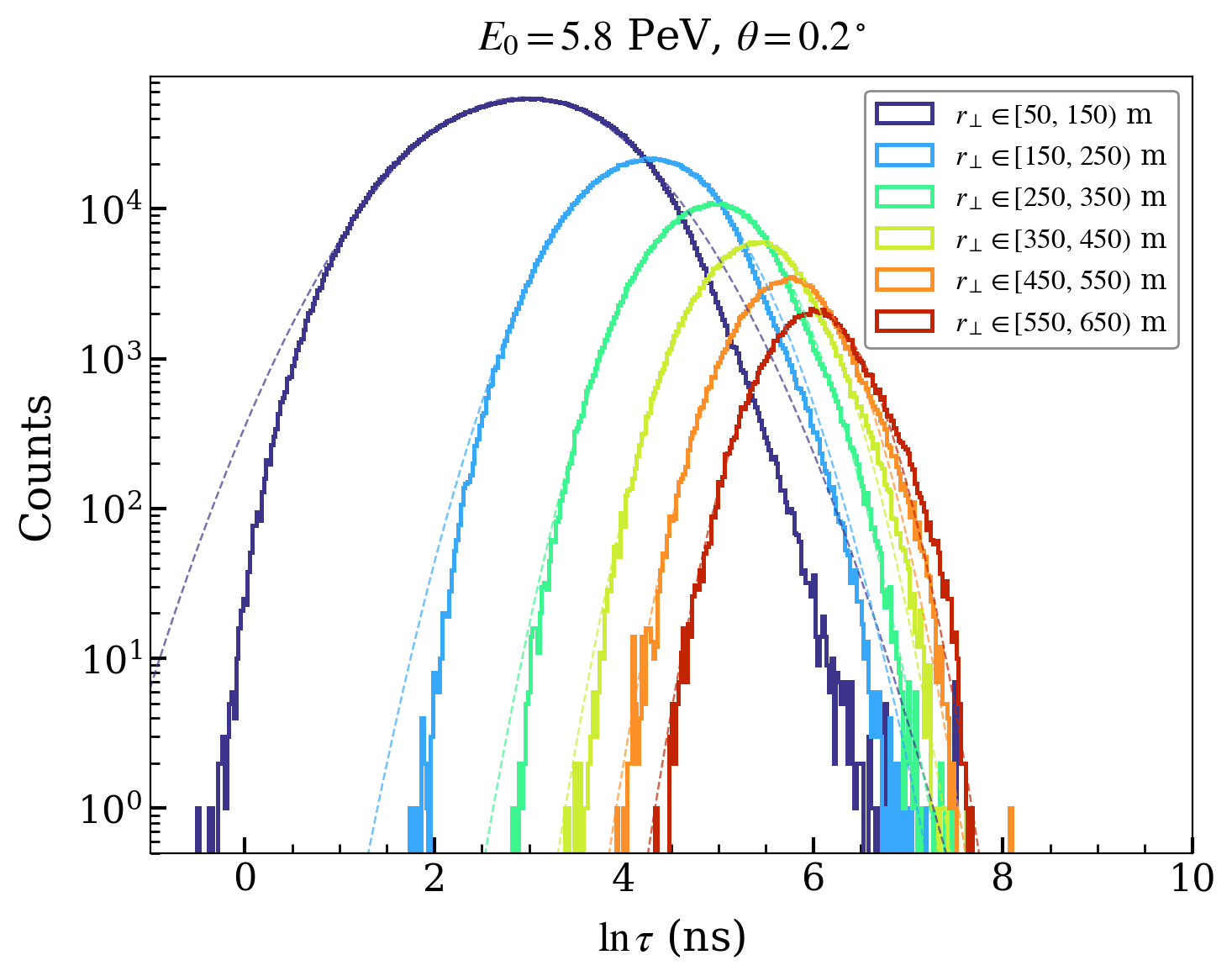}
		\caption{Distributions of $\ln\tau$ (with $\tau$ in $\mathrm{ns}$) for electromagnetic SPs from the same $\gamma$-induced air shower shown in Figure~\ref{fig:shape}. SPs within $r_\perp$ from $50$ to $650~\mathrm{m}$ are grouped into six concentric radial bins of width $\Delta r_\perp = 100~\mathrm{m}$. Each colored step curve shows the distribution of $\ln\tau$ in a given radial bin, and the dashed curve of the same color represents a Gaussian fit of the form $A\exp\left[-(\ln\tau-\mu)^2/(2\sigma^2)\right]$, corresponding to a log-normal distribution in $\tau$.}
		\label{fig:event4_lnt}
	\end{figure}

	Unlike the conical and parabolic fits, which parameterize only the geometric surface of the shower front, the log-normal timing likelihood explicitly models the statistical distribution of $\tau$. Using the same $\gamma$-induced air shower as in Figure~\ref{fig:shape}, we examine $\tau$ for SPs grouped into radial rings of fixed width in $r_\perp$. Within each ring, the distribution of $\ln\tau$ (with $\tau$ expressed in nanoseconds) is well described by a Gaussian (see Figure~\ref{fig:event4_lnt}), consistent with previous studies \citep{Lafebre2009,Bahmanabadi2018,Hofmann2020}. The corresponding log-normal probability density function is
	\begin{equation}
		f(\tau\mid r_{\perp}) = \frac{1}{\tau\,\sigma(r_{\perp})\sqrt{2\pi}}
		\exp\left[-\frac{\left(\ln \tau - \mu(r_{\perp})\right)^2}{2\sigma^2(r_{\perp})}\right].
	\end{equation}
	
\begin{figure}
		\centering
		\includegraphics[width=0.45\linewidth]{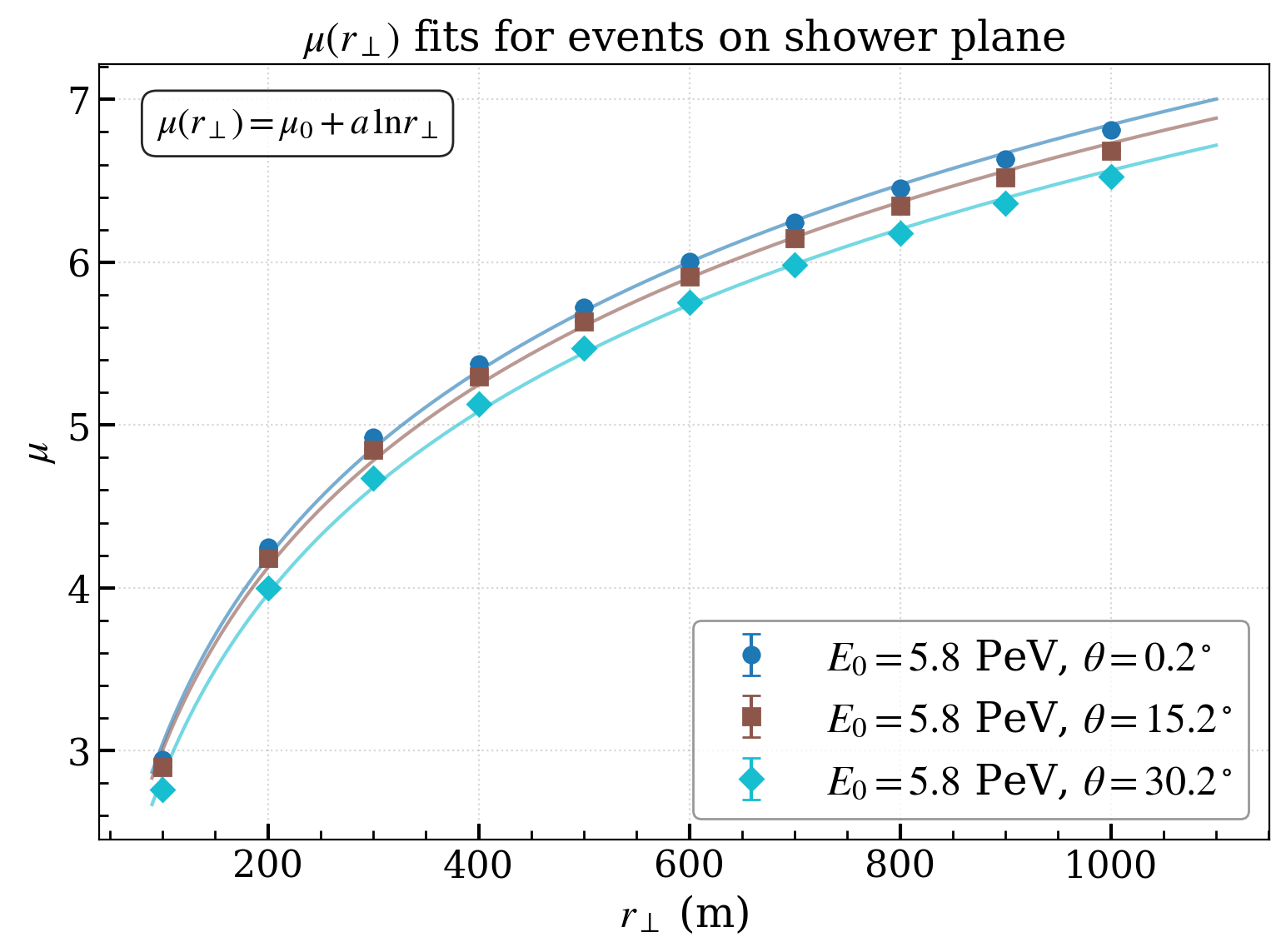}
		\hspace{-0.5ex}
		\includegraphics[width=0.45\linewidth]{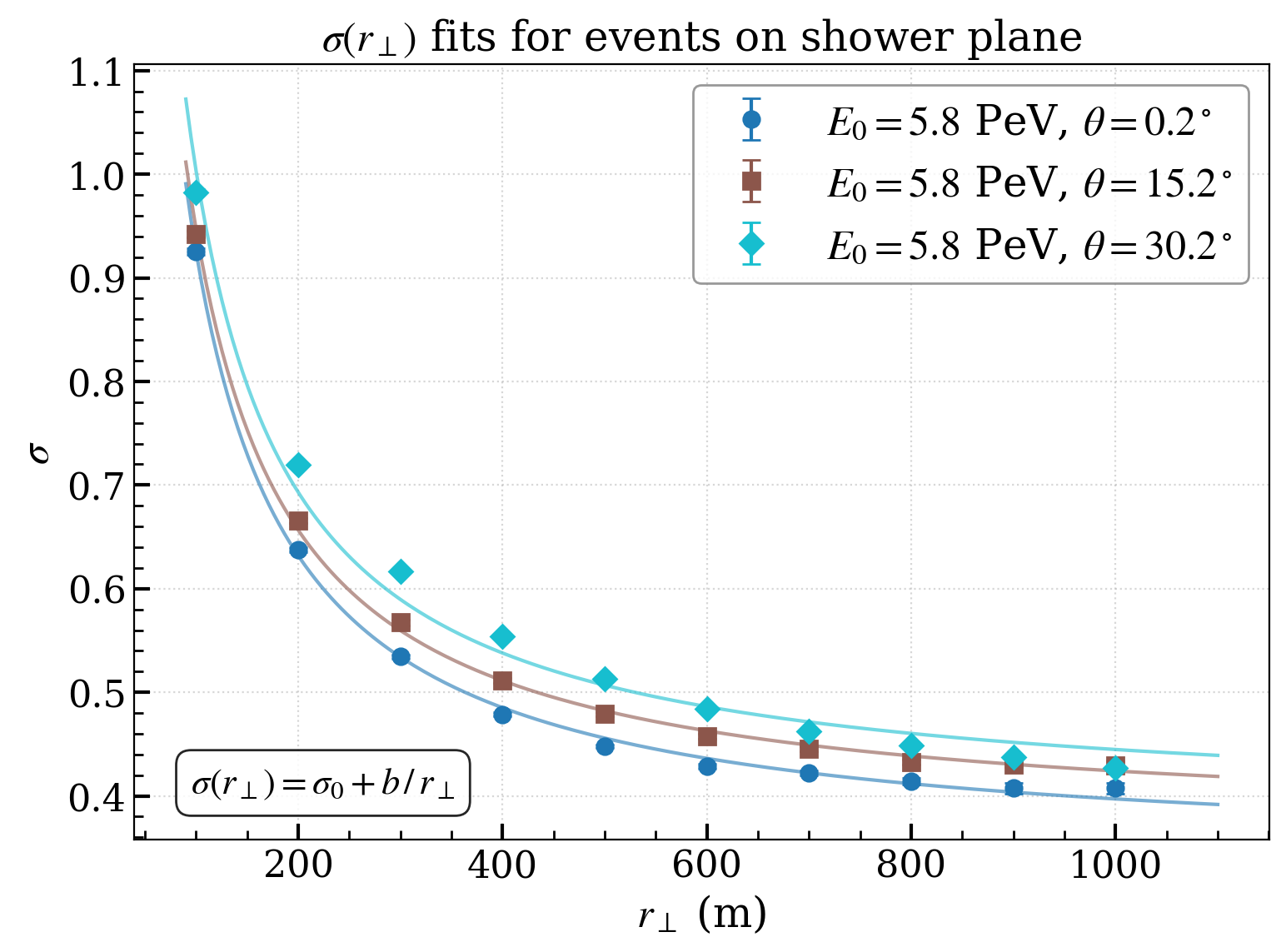}
		\\[1mm]
		\vspace{-0.8ex}
		\includegraphics[width=0.45\linewidth]{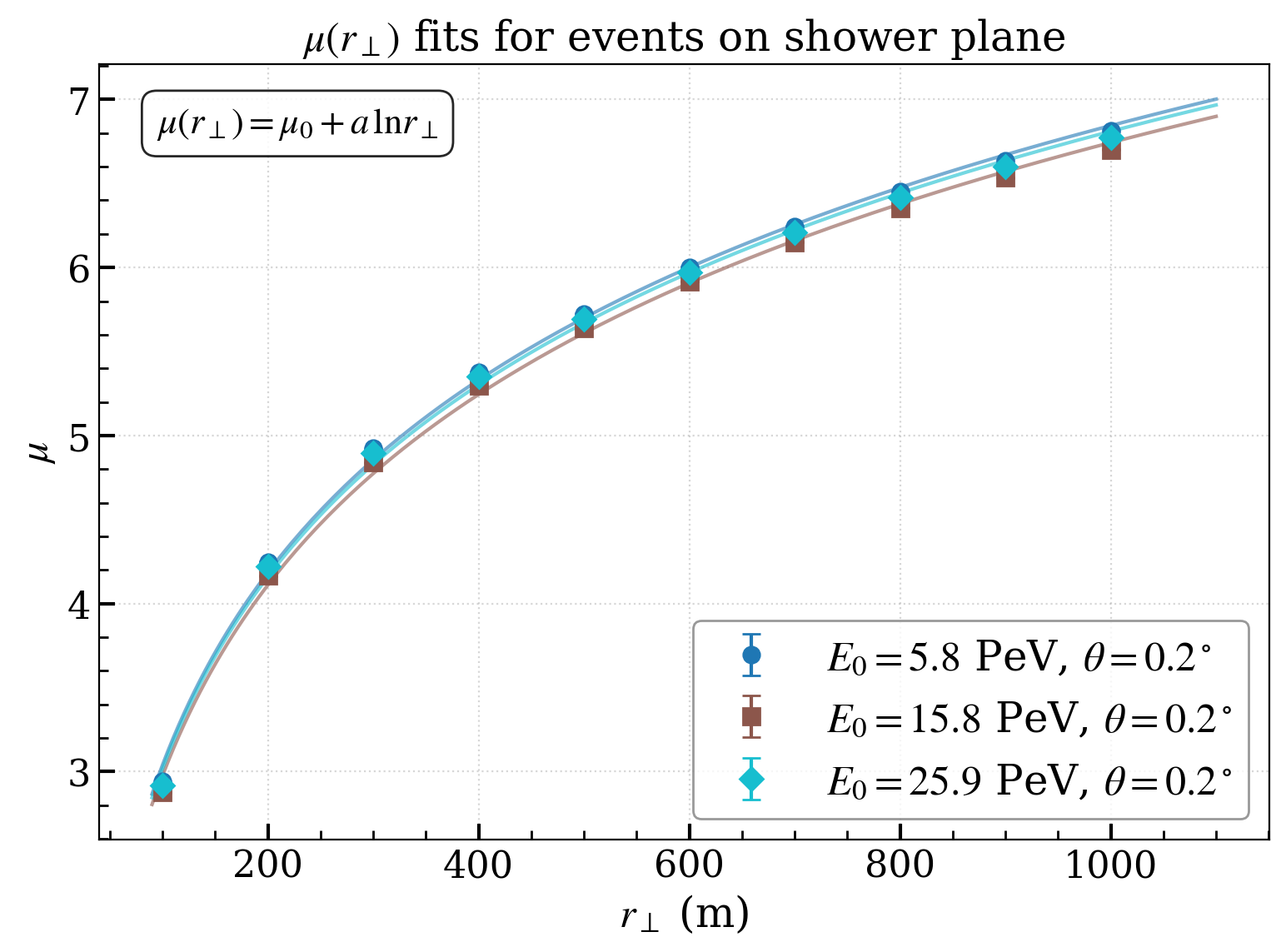}
		\hspace{-0.5ex}
		\includegraphics[width=0.45\linewidth]{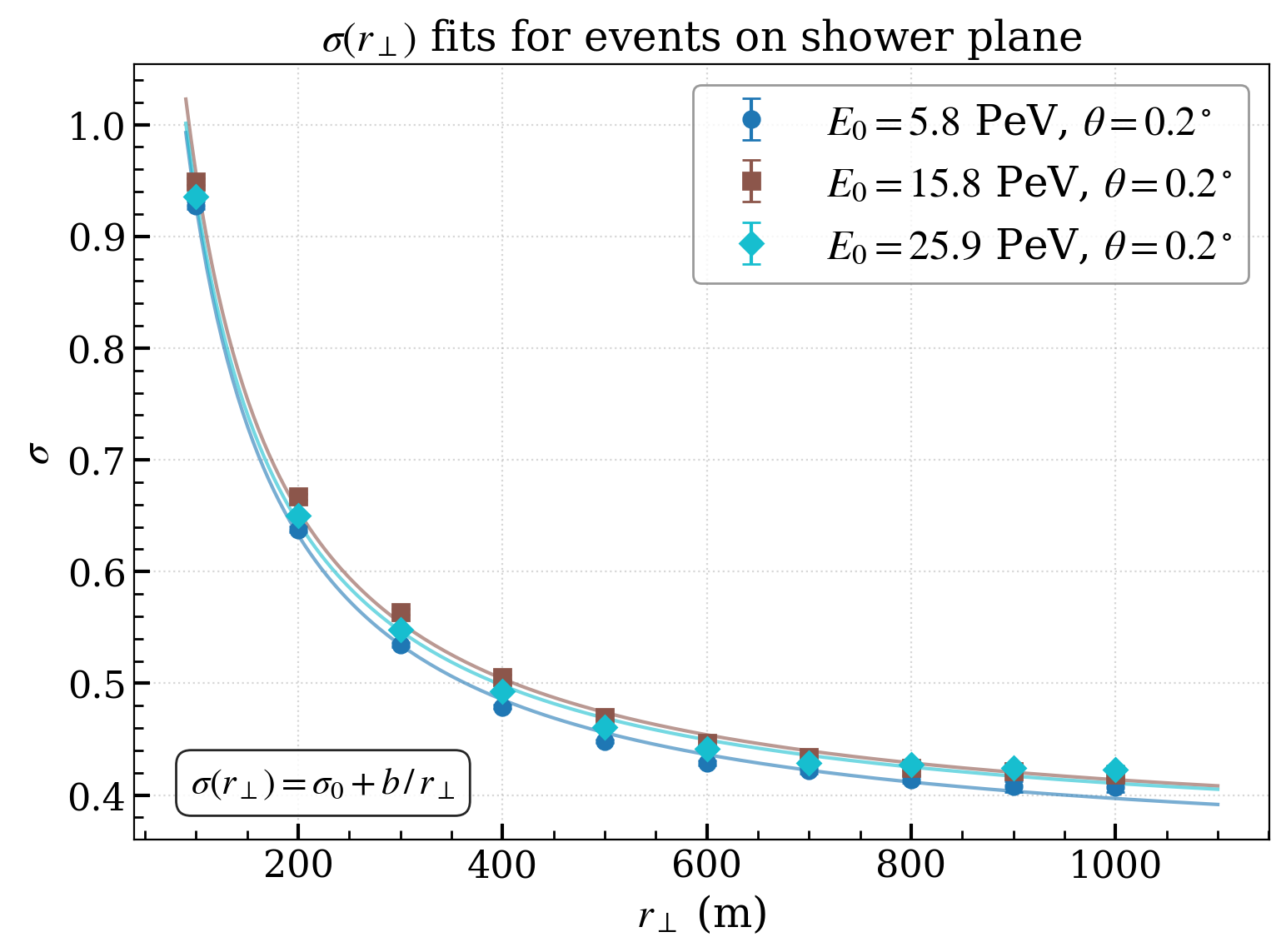}
		\caption{Radial dependence of the log-normal parameters $\mu(r_\perp)$ and $\sigma(r_\perp)$ for five $\gamma$-induced air showers. The top panels show three events with the same primary energy ($E_0 = 5.8~\mathrm{PeV}$) and different zenith angles ($\theta = 0.2^\circ, 15.2^\circ, 30.2^\circ$), while the bottom panels display three events with the same zenith angle ($\theta = 0.2^\circ$) and different primary energies ($E_0 = 5.8, 15.8, 25.9~\mathrm{PeV}$). Data points indicate the values extracted from Gaussian fits to $\ln\tau$ in individual radial rings (see Figure~\ref{fig:event4_lnt}), and solid lines represent the best-fit curves of the form $\mu(r_\perp) = \mu_0 + a\,\ln r_\perp$ (left panels) and $\sigma(r_\perp) = \sigma_0 + b/r_\perp$ (right panels), with $r_\perp$ in meters.}
		\label{fig:musigma0}
	\end{figure}

	The parameters $\mu(r_\perp)$ and $\sigma(r_\perp)$ extracted from the Gaussian fits exhibit clear empirical radial scalings:
	\begin{equation}
		\label{eq:mu_sigma_scaling}
		\mu(r_\perp) = \mu_0 + a\,\ln r_\perp,\qquad
		\sigma(r_\perp) = \sigma_0 + \frac{b}{r_\perp},
	\end{equation}
	where $r_\perp$ is expressed in meters and $(\mu_0, \sigma_0, a, b)$ are event-level model parameters. Figure~\ref{fig:musigma0} shows fits of these two relations to five representative showers: three with the same primary energy but different zenith angles, and three with the same zenith angle but different primary energies. With these scalings, the distribution of $\tau$ at any $r_\perp$ is fully specified by $(\mu_0, \sigma_0, a, b)$, forming the basis for the likelihood reconstruction described below.
	
	\subsubsection{Likelihood Implementation}
	
	For the log-normal model, Eq.~\eqref{eq:general_residual} is evaluated with $\eta = \kappa$ and $f(r_{\perp,i}) = r_{\perp,i}$, yielding
	\begin{equation}\label{eq:dti}
		\delta t_i = t_i - t_0 - \frac{1}{c}\left(\vec{r}_i \cdot \hat{n}
		+ \kappa\, r_{\perp,i}\right),
	\end{equation}
	where $\kappa$ is a free parameter fitted together with $\hat{n}$.\footnote{$\kappa$ is not a physical parameter describing the shower-front curvature; rather, it serves as an empirical correction that compensates for a residual linear radial dependence in the timing residuals and improves the direction reconstruction performance.} Unlike in the conical and parabolic fits, $t_0$ is not treated as a free parameter. Instead, for each trial direction $\hat{n}$, $t_0$ is computed directly from the data as the 2nd percentile of the positive projected arrival times in the shower plane, defining the zero point of the timing residuals. Only units with $\delta t_i > 0$ are retained in the likelihood evaluation, as required by the domain of the log-normal distribution.
	
	For a scintillator unit triggered by a single SP, including all triggered units with $0.5 < N \le 1$, we set $N = 1$ and model the probability density of $\delta t_i$ (in nanoseconds) at a given $r_\perp$ (in meters) using a log-normal distribution:
	\begin{equation}
		\label{eq:lognormal_pdf}
		f(\delta t_{i}\mid r_\perp) = \frac{1}{\delta t_{i}\,\sigma(r_\perp)\sqrt{2\pi}}
		\exp\!\left[-\frac{\bigl(\ln \delta t_{i} - \mu(r_\perp)\bigr)^2}
		{2\sigma^2(r_\perp)}\right],
	\end{equation}
	where $\mu(r_\perp)$ and $\sigma(r_\perp)$ follow Eq.~\eqref{eq:mu_sigma_scaling}. The likelihood contribution of this unit is $\mathcal{L}_i(\delta t_{i}\mid r_\perp) = f(\delta t_{i}\mid r_\perp)$.
	
	When a scintillator unit records a signal with effective multiplicity $N > 1$, the measured trigger time corresponds to the earliest arrival among the contributing SPs. Assuming that their arrival times are independent draws from the same local log-normal distribution at $r_\perp$, the probability density for this earliest arrival is \citep{David2003}
	\begin{equation}
		\label{eq:station_likelihood}
		\mathcal{L}_i(\delta t_{i}\mid r_\perp) = N\,f(\delta t_{i}\mid r_\perp)\,\bigl[S(\delta t_{i}\mid r_\perp)\bigr]^{N-1}.
	\end{equation}
	Here, $S(\delta t_{i}\mid r_\perp) = 1 - F(\delta t_{i}\mid r_\perp)$ is the survival function, and $F(\delta t_{i}\mid r_\perp)$ is the cumulative distribution function of the log-normal distribution:
	\begin{equation}
		\label{eq:lognormal_cdf}
		F(\delta t_{i}\mid r_\perp) = \frac{1}{2}\left[1 + \mathrm{erf}\!\left(\frac{\ln \delta t_{i} - \mu(r_\perp)}{\sqrt{2}\,\sigma(r_\perp)}\right)\right].
	\end{equation}
	
\begin{figure}
		\centering
		\includegraphics[width=0.65\linewidth]{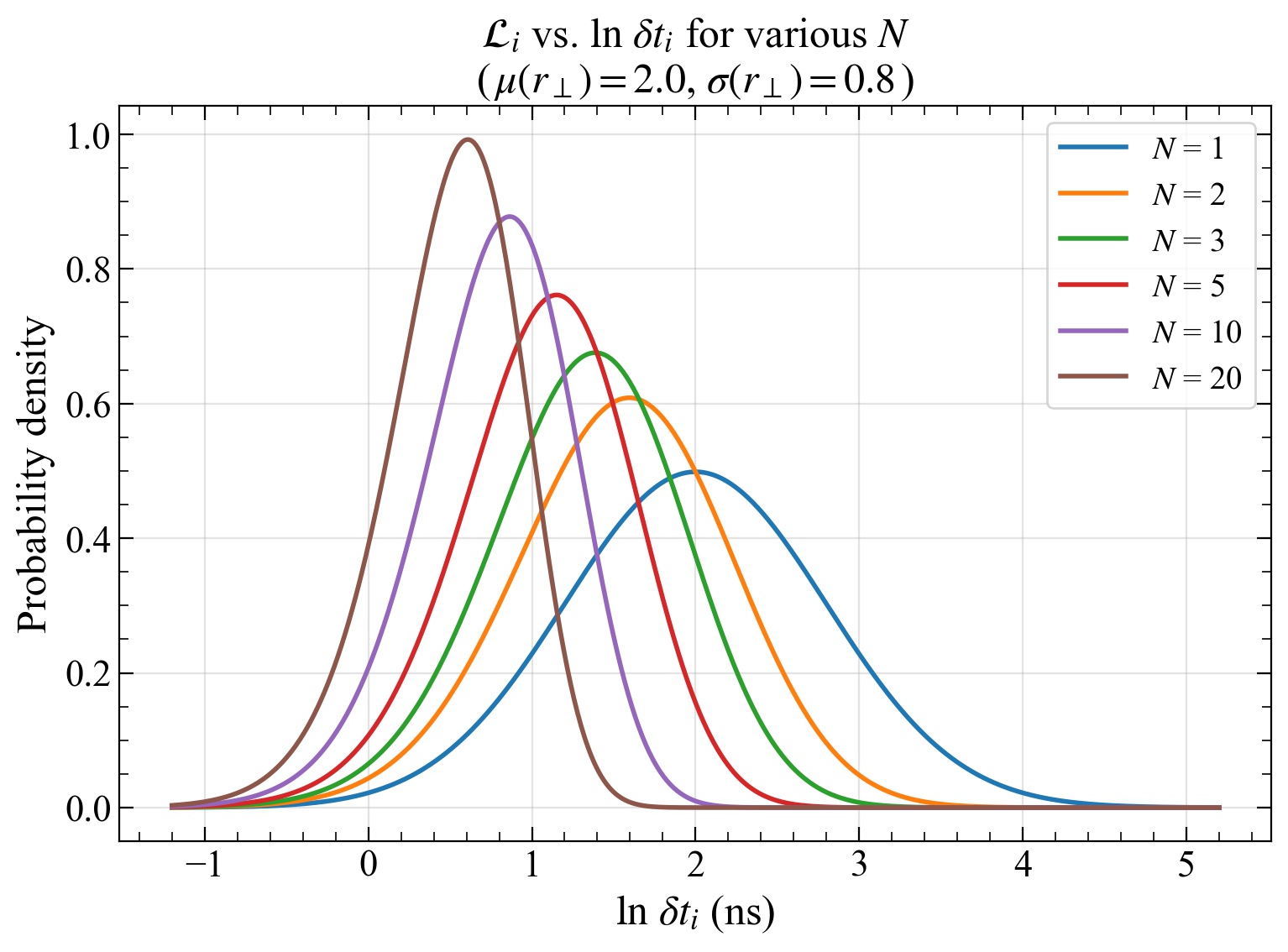}
		\caption{Likelihood contribution $\mathcal{L}_i$ of a single detector unit as a function of $\ln\delta t_{i}$ (with $\delta t_i$ in $\mathrm{ns}$) for several values of the effective multiplicity $N$, with log-normal parameters fixed at $\mu(r_\perp) = 2.0$ and $\sigma(r_\perp) = 0.8$.}
		\label{fig:N_dependence}
	\end{figure}

	To illustrate the behavior of $\mathcal{L}_i$, Figure~\ref{fig:N_dependence} shows $\mathcal{L}_i$ as a function of $\ln\delta t_{i}$ for several values of $N$, with $\mu(r_\perp) = 2.0$ and $\sigma(r_\perp) = 0.8$. For $N = 1$, $\mathcal{L}_i$ reduces to the single-particle log-normal distribution. As $N$ increases, the distribution narrows, its peak shifts toward smaller $\ln\delta t_{i}$, and its maximum density rises, reflecting the higher probability of an earlier trigger when more particles contribute to the signal.
	
	Combining all triggered units, the objective function for an event is the negative log-likelihood,
	\begin{equation}
		\label{eq:total_loglikelihood}
		\mathcal{F} = \sum_i \rho_i(\delta t_i) = -\sum_i \ln \mathcal{L}_i(\delta t_i\mid r_{\perp,i}).
	\end{equation}
	For each trial pair $(\hat{n}, \kappa)$, after updating $t_0$, the triggered units are grouped into radial rings to extract $\mu(r_\perp)$ and $\sigma(r_\perp)$ and fit the four coefficients $(\mu_0, \sigma_0, a, b)$. Because each radial scaling relation in Eq.~\eqref{eq:mu_sigma_scaling} involves only two parameters, grouping the measurements into at least two to three radial rings is sufficient to determine these coefficients even for events near the $N_{\rm TrigE} > 9$ threshold. The objective function $\mathcal{F}$ is then minimized with respect to $(\hat{n}, \kappa)$ using the \textsc{MINUIT} package \citep{minuit} to obtain the reconstructed shower direction.
	
	\section{Results}
	
	\subsection{Angular Resolution as a Function of Triggered Stations}
	\label{sec:res_ntrige}
	
\begin{figure}
		\centering
		\includegraphics[width=0.7\linewidth]{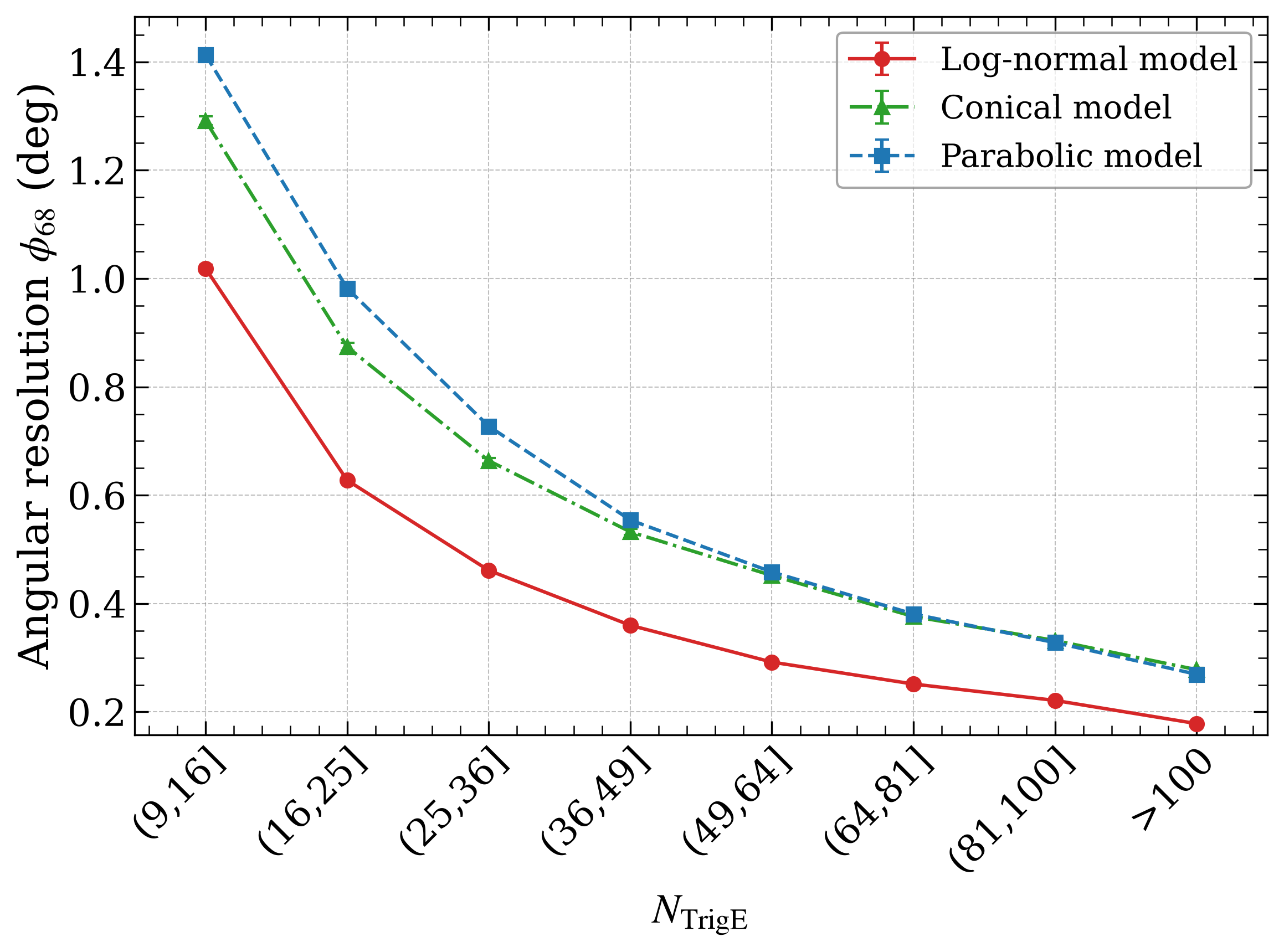}
		\caption{Angular resolution (68\% containment $\phi_{68}$) as a function of the number of triggered detector stations, $N_{\mathrm{TrigE}}$, for the log-normal model (red circles), the conical model (green triangles), and the parabolic model (blue squares). Error bars denote 68\% bootstrap confidence intervals.}
		\label{fig:res_ntrige}
	\end{figure}

	We first compare the angular resolution of the three reconstruction models as a function of the number of triggered stations, $N_{\mathrm{TrigE}}$. Figure~\ref{fig:res_ntrige} shows the 68\% containment angle $\phi_{68}$ for each method, and Table~\ref{tab:res_ntrige} lists the corresponding numerical values. In all three cases, $\phi_{68}$ improves monotonically with $N_{\mathrm{TrigE}}$ as more spatially distributed timing measurements constrain the shower axis. Across the full $N_{\mathrm{TrigE}}$ range, the log-normal likelihood outperforms both the conical and parabolic wavefront fits.
	
	This difference reflects how each method handles arrival-time fluctuations. Because individual detector units sample a finite number of SPs, measured trigger times fluctuate around the nominal shower front. These fluctuations grow at large $r_{\perp}$, where the lateral particle density drops \citep{Kamata1958,Greisen1960,Miyake1961} and the shower front thickens \citep{Inoue1985,Linsley1986,Agnetta1997,Huang2019}, producing a broad and strongly skewed arrival-time distribution at far-axis stations.
	
	\begin{table}[H]
		\centering
		\caption{Angular resolution (68\% containment $\phi_{68}$) in different $N_{\mathrm{TrigE}}$ intervals for the three reconstruction models. Uncertainties are statistical only, obtained from 500 bootstrap re-samples.\protect\footnotemark~ Since the point-source sensitivity of an EAS array improves inversely with the angular resolution, the last column lists the resolution ratios, $\phi_{68}^{\rm Con}/\phi_{68}^{\rm Log}$ and $\phi_{68}^{\rm Par}/\phi_{68}^{\rm Log}$, serving as a direct proxy for the sensitivity gain provided by the log-normal likelihood.}
		\label{tab:res_ntrige}\vspace{+0.9ex}
		\begin{tabular}{|ccccc|}
			\hline
			$N_{\mathrm{TrigE}}$ & Log-normal & Conical & Parabolic & Resolution ratio \\
			& (deg) & (deg) & (deg) & ($\phi^{\rm Con}/\phi^{\rm Log}$ \,\&\, $\phi^{\rm Par}/\phi^{\rm Log}$) \\
			\hline
			$(9,16]$    & $1.019\pm0.007$ & $1.291\pm0.008$ & $1.413\pm0.010$ & $1.27$ \& $1.39$ \\
			$(16,25]$   & $0.628\pm0.005$ & $0.874\pm0.007$ & $0.982\pm0.007$ & $1.39$ \& $1.56$ \\
			$(25,36]$   & $0.461\pm0.004$ & $0.664\pm0.005$ & $0.728\pm0.006$ & $1.44$ \& $1.58$ \\
			$(36,49]$   & $0.360\pm0.004$ & $0.532\pm0.005$ & $0.554\pm0.004$ & $1.48$ \& $1.54$ \\
			$(49,64]$   & $0.292\pm0.004$ & $0.452\pm0.006$ & $0.458\pm0.006$ & $1.55$ \& $1.57$ \\
			$(64,81]$   & $0.251\pm0.004$ & $0.376\pm0.005$ & $0.381\pm0.006$ & $1.50$ \& $1.52$ \\
			$(81,100]$  & $0.221\pm0.003$ & $0.332\pm0.005$ & $0.328\pm0.005$ & $1.50$ \& $1.48$ \\
			$>100$      & $0.178\pm0.003$ & $0.278\pm0.004$ & $0.269\pm0.004$ & $1.56$ \& $1.51$ \\
			\hline
		\end{tabular}
	\end{table}
	\footnotetext{%
		For each $N_{\mathrm{TrigE}}$ interval, we draw 500 resamples with
		replacement from the angular errors between the reconstructed and true
		directions in that interval, compute the 68\% containment angle
		$\phi_{68}$ for each resample, and quote the standard deviation of these 500
		estimates as the uncertainty. This bootstrap estimate approximately follows
		the asymptotic scaling $\sigma_{68}\approx 0.64\,\phi_{68}/\sqrt{N}$, where
		$N$ is the number of events in the interval and the coefficient 0.64 follows
		from the Rayleigh distribution of radial errors for a two-dimensional
		Gaussian \citep{Efron1993}.}
	
	Conical and parabolic fits suppress these fluctuations through spatial and temporal weights that down-weight detectors at large $r_\perp$ and penalize early or late outliers. While this limits the pull of extreme residuals, it also discards useful directional information from far-axis detectors. By contrast, the log-normal likelihood models the asymmetric arrival-time distribution directly and accounts for multiple particles hitting a single unit. Retaining the statistical information in the non-Gaussian tails improves the angular resolution across all $N_{\mathrm{TrigE}}$ intervals. As shown by the resolution ratios in Table~\ref{tab:res_ntrige}, this corresponds to a potential point-source sensitivity gain of up to a factor of $\sim 1.5$.
	
	\subsection{Impact of Intra-Station Spatial Sampling on Angular Resolution}
	\label{sec:res_density}
	
	We next examine how intra-station spatial sampling affects the angular resolution by comparing two readout schemes on the same simulated sample. Under the \emph{per-unit} scheme, each of the four scintillator units in a station is read out independently with a $5~\mathrm{MeV}$ threshold. Under the \emph{per-station} scheme, the four units are merged into a single detector whose deposited energy is the sum of the four unit deposits, subject to the same $5~\mathrm{MeV}$ threshold, with the trigger time taken as the earliest unit trigger and the position set to the station center. We apply the log-normal reconstruction independently to both data sets.
	
\begin{figure}[htbp]
		\centering
		\includegraphics[width=0.7\textwidth]{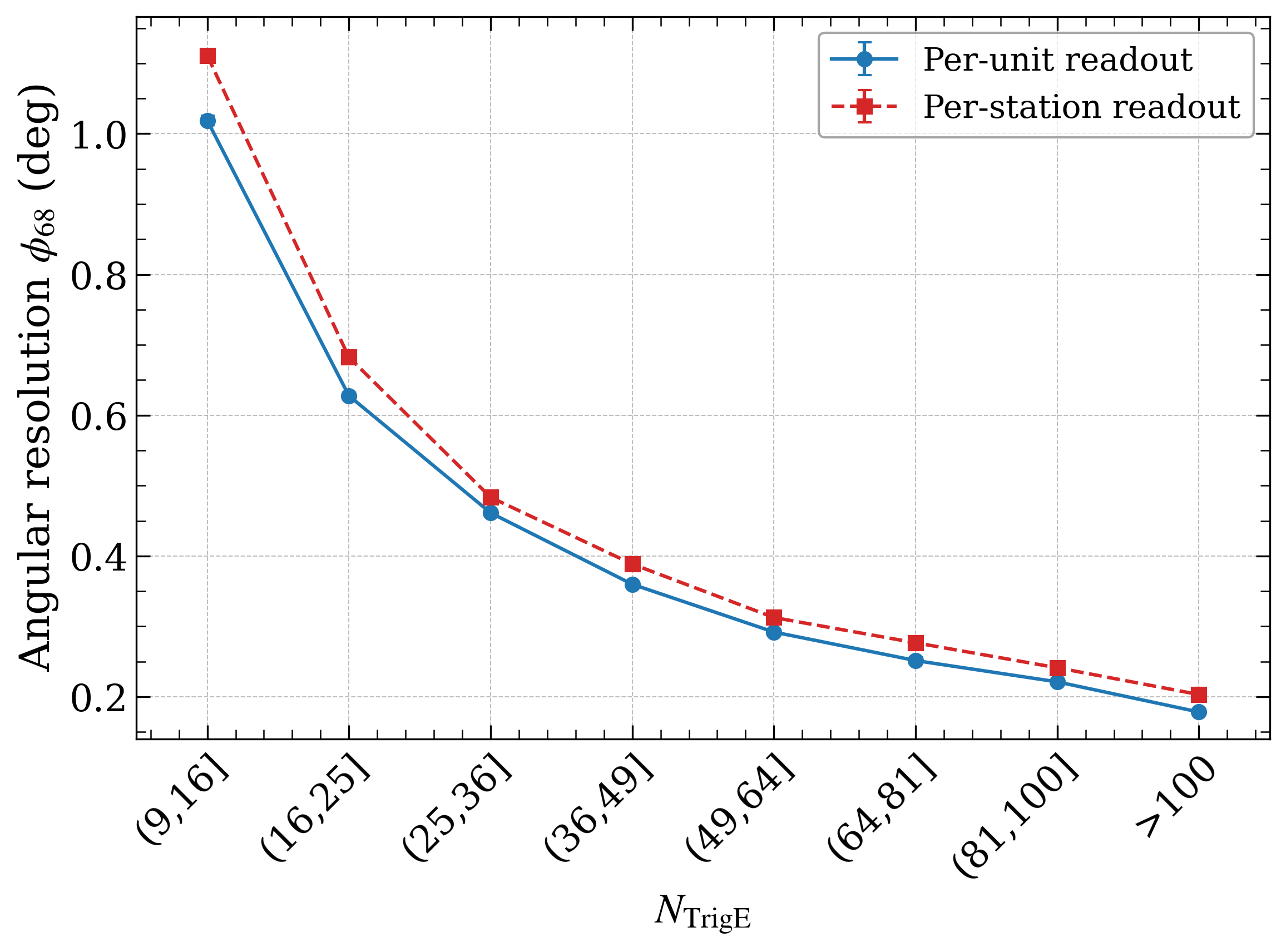}
		\caption{Angular resolution at 68\% containment for the log-normal model as a function of the number of triggered stations $N_{\mathrm{TrigE}}$ under two readout schemes: per-unit readout (blue circles), in which each unit contributes its own measurement to the likelihood, and per-station readout (red squares), in which the four units in a station are merged into a single station-level detector. $N_{\mathrm{TrigE}}$ is defined according to each scheme's own trigger condition. Error bars denote 68\% bootstrap confidence intervals.}
		\label{fig:res_station}
	\end{figure}

	Figure~\ref{fig:res_station} and Table~\ref{tab:res_density} compare the resulting angular resolutions. The per-unit readout performs better across the full $N_{\mathrm{TrigE}}$ range because it retains redundant local timing measurements. Rather than reducing a station to a single trigger time, each triggered unit enters the likelihood as a separate constraint, reducing the impact of single-unit timing fluctuations on the global fit.
	
	The relative gain shows a non-monotonic dependence on $N_{\mathrm{TrigE}}$, which likely reflects the balance between global shower-front constraints and local intra-station redundancy. At low $N_{\mathrm{TrigE}}$, the global fit is sensitive to individual timing fluctuations, and keeping per-unit measurements appears to stabilize these local terms, giving the largest absolute improvement. At intermediate $N_{\mathrm{TrigE}}$, the larger number of triggered stations may already average out isolated anomalies, reducing the marginal gain from intra-station redundancy. At high $N_{\mathrm{TrigE}}$, near-core stations experience high particle densities and typically trigger all four units; quadrupling the timing constraints in this well-sampled region likely accounts for the renewed rise in relative improvement.
	
	\begin{table}[htbp]
		\centering
		\caption{Angular resolution (68\% containment $\phi_{68}$) obtained with the log-normal model as a function of $N_{\mathrm{TrigE}}$, comparing the per-station readout (four units merged into one station-level detector) with the per-unit readout (each unit contributing its own measurement to the likelihood). Uncertainties are statistical only, obtained from 500 bootstrap re-samples, as described in the footnote of Table~\ref{tab:res_ntrige}. The last column lists the relative improvement $(\phi_{68}^{\text{per-station}}-\phi_{68}^{\text{per-unit}})/\phi_{68}^{\text{per-station}}$, expressed in percent.}
		\label{tab:res_density}\vspace{+0.9ex}
		\begin{tabular}{|cccc|}
			\hline
			$N_{\mathrm{TrigE}}$ & Per-station & Per-unit & Relative improvement \\
			& (deg) & (deg) & (\%) \\
			\hline
			$(9,16]$    & $1.111\pm0.007$ & $1.019\pm0.007$ & $8.3$ \\
			$(16,25]$   & $0.682\pm0.005$ & $0.628\pm0.005$ & $7.9$ \\
			$(25,36]$   & $0.483\pm0.004$ & $0.461\pm0.004$ & $4.6$ \\
			$(36,49]$   & $0.389\pm0.004$ & $0.360\pm0.004$ & $7.5$ \\
			$(49,64]$   & $0.313\pm0.004$ & $0.292\pm0.004$ & $6.7$ \\
			$(64,81]$   & $0.277\pm0.004$ & $0.251\pm0.004$ & $9.4$ \\
			$(81,100]$  & $0.241\pm0.004$ & $0.221\pm0.003$ & $8.3$ \\
			$>100$      & $0.203\pm0.003$ & $0.178\pm0.003$ & $12.3$ \\
			\hline
		\end{tabular}
	\end{table}
	
	\section{Conclusions and Discussion}
	
	In this work, we developed a log-normal timing likelihood for reconstructing the arrival direction of UHE $\gamma$-ray air showers over large lateral distances. Using Monte Carlo simulations of a sparse surface array, we compared this approach with conventional conical and parabolic wavefront fits. The log-normal likelihood achieves better angular resolution across the entire range of triggered station multiplicities. This improvement arises from explicitly modeling the positively skewed arrival-time distribution of SPs in the shower plane and accounting for multiple particles triggering a single detector unit. By using the full timing information from far-axis detectors rather than down-weighting non-Gaussian tails, the method provides a more accurate direction estimate than standard $\chi^2$ fits.
	
	This gain is especially relevant for future large-area observatories. In our simulations, the relative reduction in angular error is largest at high triggered multiplicities, corresponding to high-energy primary $\gamma$ rays that illuminate detectors over wide lateral distances. At these distances, far-axis arrival-time distributions become strongly asymmetric, causing weighted least-squares fits to lose directional information. Because the log-normal likelihood naturally accommodates these broad, skewed distributions, it maintains a clear advantage even as the absolute angular error approaches the instrumental timing floor.
	
	We also found that reading out individual detector units independently improves the angular resolution over merging them into a single station-level signal. Treating each triggered unit as a separate timing constraint increases local sampling redundancy and reduces the influence of isolated timing fluctuations. The relative benefit is largest at high multiplicities, where high near-core particle densities trigger all four units in a station and provide dense local timing measurements. Finer intra-station granularity therefore offers a straightforward way to further improve angular resolution in sparse arrays.
	
	One limitation of this study is the use of the true shower core position. In an actual experiment, core reconstruction uncertainties will propagate into the lateral distance estimates and degrade the angular resolution. Because the log-normal likelihood captures the radial evolution of the shower front, however, it can be extended directly into a joint fit for both the shower core and the arrival direction. In addition, the shape parameters of the timing distribution may carry information about longitudinal shower development, particularly the depth of shower maximum ($X_{\max}$), offering a potential observable for characterizing the primary particle. With dominant instrumental timing limits already included in our simulations, the log-normal timing likelihood provides a transparent, physically interpretable, and robust reconstruction framework for future UHE $\gamma$-ray observatories.
	
	\acknowledgments
	This research work is supported by the following grants: the National SKA Program of China (No. 2025SKA0110103), the National Natural Science Foundation of China (Nos. 12220101003, 12273114), the Project for Young Scientists in Basic Research of the Chinese Academy of Sciences (No. YSBR-061), and Jiangsu Innovation and Entrepreneurship Talent Team Program (No. JSSCTD202436).
	
	\bibliography{biblio}{}

@article{LHAASO2026PSRJ1849,
  author       = {Cao, Zhen and Aharonian, Felix and Bai, Yun-Xiang and Bao, Yi-Wei and Bastieri, Denis and others},
  title        = {An extreme particle accelerator powered by pulsar PSR J1849−0001},
  journal      = {Nature Astronomy},
  year         = {2026},
  volume       = {10},
  pages        = {1029--1038},
  doi          = {10.1038/s41550-026-02839-0},
  eprint       = {2603.15537},
  archivePrefix= {arXiv},
  primaryClass = {astro-ph.HE}
}

@article{LHAASO2026Peanut,
  author       = {Cao, Zhen and Aharonian, Felix and Bai, Yun-Xiang and Bao, Yi-Wei and Bastieri, Denis and others},
  title        = {A Giant Peanut-shaped Ultra-High-Energy Gamma-Ray Emitter Off the Galactic Plane},
  journal      = {Science Bulletin},
  year         = {2026},
  doi          = {10.1016/j.scib.2026.09.046},
  eprint       = {2510.0678},
  archivePrefix= {arXiv},
  primaryClass = {astro-ph.HE}
}

@article{LHAASO2026CygnusX3,
  author       = {Cao, Zhen and Aharonian, Felix and Bai, Yun-Xiang and Bao, Yi-Wei and Bastieri, Denis and others},
  title        = {Cygnus {X-3}: A variable petaelectronvolt $\gamma$-ray source},
  journal      = {National Science Review},
  year         = {2026},
  volume       = {nwag435},
  doi          = {10.1093/nsr/nwag435},
  eprint       = {2512.16638},
  archivePrefix= {arXiv},
  primaryClass = {astro-ph.HE}
}

@misc{TibetConeFit,
    author  = {Okukawa, S. and Kobayashi, M. and Hara, K. and
                     Hibino, K. and Katayose, Y. and Kawata, K. and
                     Ohnishi, M. and Sako, T. K. and Sako, T. and
                     Shiomi, A. and Takita, M.},
    title         = {A Deep Learning Approach to Improving the
                     Arrival-Direction Accuracy of Gamma-Ray
                     Air-Shower Measurements with the Tibet-III
                     Air-Shower Array},
    year          = {2026},
    eprint        = {2610.09736},
    archivePrefix = {arXiv},
    primaryClass  = {astro-ph.IM}
}

@article{HAWCPass5,
  author  = {Albert, A. and Alfaro, R. and Alvarez, C. and others},
  title   = {Performance of the {HAWC} {Observatory} and {TeV} Gamma-Ray Measurements of the {Crab} Nebula with Improved Extensive Air Shower Reconstruction Algorithms},
  journal = {The Astrophysical Journal},
  volume  = {972},
  pages   = {144},
  year    = {2024},
  doi     = {10.3847/1538-4357/ad5f2d}
}

@phdthesis{Milagro,
    author = {Benbow, Wystan},
    title  = {A Study of TeV Emission from the Crab Nebula and Selected AGN Using the Milagro Gamma-Ray Observatory},
    school = {University of California, Santa Cruz},
    year   = {2002}
}

@article{Atkins2004,
    author  = {Atkins, R. and others},
    title   = {Search for Very High Energy Gamma Rays from WIMP
               Annihilations near the Sun with the Milagro Detector},
    journal = {Physical Review D},
    volume  = {70},
    pages   = {083516},
    year    = {2004},
    doi     = {10.1103/PhysRevD.70.083516}
}

@phdthesis{KASCADEGrande,
    author = {Łuczak, Paweł},
    title  = {Investigation of Muons in Extensive Air Showers with the KASCADE-Grande Muon Tracking Detector},
    school = {National Centre for Nuclear Research},
    year   = {2012}
}

@article{Antoni2004KASCADE,
    author  = {Antoni, T. and others},
    title   = {Search for Cosmic-Ray Point Sources with {KASCADE}},
    journal = {The Astrophysical Journal},
    volume  = {608},
    pages   = {865--871},
    year    = {2004},
    doi     = {10.1086/420736}
}

@book{David2003,
    author    = {David, H. A. and Nagaraja, H. N.},
    title     = {Order Statistics},
    edition   = {3},
    publisher = {John Wiley \& Sons},
    address   = {Hoboken, New Jersey},
    year      = {2003},
    isbn      = {978-0-471-38926-2},
    doi       = {10.1002/0471722162}
}

@book{Efron1993,
  author    = {Efron, Bradley and Tibshirani, Robert J.},
  title     = {An Introduction to the Bootstrap},
  publisher = {Springer},
  year      = {1993},
  doi       = {10.1007/978-1-4899-4541-9},
  isbn      = {978-1-4899-4541-9}
}

@TECHREPORT{Heck1998,
    author      = {D. Heck and J. Knapp and J. N. Capdevielle and G. Schatz and T. Thouw},
    title       = {{CORSIKA}: A Monte Carlo Code to Simulate Extensive Air Showers},
    institution = {Forschungszentrum Karlsruhe},
    number      = {FZKA 6019},
    year        = {1998},
    doi         = {10.5445/IR/270043064}
}

@ARTICLE{Riehn2020,
    author  = {F. Riehn and R. Engel and A. Fedynitch and T. K. Gaisser and T. Stanev},
    title   = {Hadronic interaction model {SIBYLL} 2.3d and extensive air showers},
    journal = {Phys. Rev. D},
    volume  = {102},
    year    = {2020},
    pages   = {063002},
    doi     = {10.1103/PhysRevD.102.063002}
}

@ARTICLE{Aab2014,
       author = {{Pierre Auger Collaboration}},
        title = "{Reconstruction of inclined air showers detected with the Pierre Auger Observatory}",
      journal = {\jcap},
         year = 2014,
        month = aug,
       volume = {2014},
       number = {8},
        pages = {019-019},
          doi = {10.1088/1475-7516/2014/08/019},
archivePrefix = {arXiv},
       eprint = {1407.3214},
 primaryClass = {astro-ph.HE},
       adsurl = {https://ui.adsabs.harvard.edu/abs/2014JCAP...08..019P}
}

@ARTICLE{Agnetta1997,
       author = {{Agnetta}, G. and {Ambrosio}, M. and {Aramo}, C. and {Barbarino}, G.~C. and {Beaman}, J. and {Biondo}, B. and {Campana}, D. and {Catalano}, O. and {Colesanti}, L. and {D'Ali}, G. and {Erlykin}, A. and {Guarino}, F. and {Lauro}, A. and {Lloyd-Evans}, J. and {Mangano}, A. and {Osteria}, G. and {Watson}, A.~A.},
        title = "{Time structure of the extensive air shower front}",
      journal = {Astroparticle Physics},
         year = 1997,
        month = mar,
       volume = {6},
       number = {3-4},
        pages = {301-312},
          doi = {10.1016/S0927-6505(96)00064-3},
       adsurl = {https://ui.adsabs.harvard.edu/abs/1997APh.....6..301A}
}

@article{geant4,
  author = {Agostinelli, S. and others},
  title = {GEANT4: a simulation toolkit},
  journal = {Nucl. Instrum. Meth. A},
  year = {2003},
  volume = {506},
  number = {3},
  pages = {250--303},
  doi = {10.1016/S0168-9002(03)01368-8},
  adsurl = {https://ui.adsabs.harvard.edu/abs/2003NIMPA.506..250A}
}

@ARTICLE{Aharonian2021,
       author = {{Aharonian}, F. and {An}, Q. and {Axikegu} and {Bai}, L.~X. and {Bai}, Y.~X. and {Bao}, Y.~W. and {Bastieri}, D. and {Bi}, X.~J. and {Bi}, Y.~J. and {Cai}, H. and {Cai}, J.~T. and {Cao}, Z. and {Cao}, Z. and {Chang}, J. and {Chang}, J.~F. and {Chang}, X.~C. and {Chen}, B.~M. and {Chen}, J. and {Chen}, L. and {Chen}, L. and {Chen}, L. and {Chen}, M.~J. and {Chen}, M.~L. and {Chen}, Q.~H. and {Chen}, S.~H. and {Chen}, S.~Z. and {Chen}, T.~L. and {Chen}, X.~L. and {Chen}, Y. and {Cheng}, N. and {Cheng}, Y.~D. and {Cui}, S.~W. and {Cui}, X.~H. and {Cui}, Y.~D. and {Dai}, B.~Z. and {Dai}, H.~L. and {Dai}, Z.~G. and {Danzengluobu} and {Della Volpe}, D. and {Piazzoli}, B. D'ettorre and {Dong}, X.~J. and {Fan}, J.~H. and {Fan}, Y.~Z. and {Fan}, Z.~X. and {Fang}, J. and {Fang}, K. and {Feng}, C.~F. and {Feng}, L. and {Feng}, S.~H. and {Feng}, Y.~L. and {Gao}, B. and {Gao}, C.~D. and {Gao}, Q. and {Gao}, W. and {Ge}, M.~M. and {Geng}, L.~S. and {Gong}, G.~H. and {Gou}, Q.~B. and {Gu}, M.~H. and {Guo}, J.~G. and {Guo}, X.~L. and {Guo}, Y.~Q. and {Guo}, Y.~Y. and {Han}, Y.~A. and {He}, H.~H. and {He}, H.~N. and {He}, J.~C. and {He}, S.~L. and {He}, X.~B. and {He}, Y. and {Heller}, M. and {Hor}, Y.~K. and {Hou}, C. and {Hou}, X. and {Hu}, H.~B. and {Hu}, S. and {Hu}, S.~C. and {Hu}, X.~J. and {Huang}, D.~H. and {Huang}, Q.~L. and {Huang}, W.~H. and {Huang}, X.~T. and {Huang}, Z.~C. and {Ji}, F. and {Ji}, X.~L. and {Jia}, H.~Y. and {Jiang}, K. and {Jiang}, Z.~J. and {Jin}, C. and {Kuleshov}, D. and {Levochkin}, K. and {Li}, B.~B. and {Li}, C. and {Li}, C. and {Li}, F. and {Li}, H.~B. and {Li}, H.~C. and {Li}, H.~Y. and {Li}, J. and {Li}, K. and {Li}, W.~L. and {Li}, X. and {Li}, X. and {Li}, X.~R. and {Li}, Y. and {Li}, Y.~Z. and {Li}, Z. and {Li}, Z. and {Liang}, E.~W. and {Liang}, Y.~F. and {Lin}, S.~J. and {Liu}, B. and {Liu}, C. and {Liu}, D. and {Liu}, H. and {Liu}, H.~D. and {Liu}, J. and {Liu}, J.~L. and {Liu}, J.~S. and {Liu}, J.~Y. and {Liu}, M.~Y. and {Liu}, R.~Y. and {Liu}, S.~M. and {Liu}, W. and {Liu}, Y.~N. and {Liu}, Z.~X. and {Long}, W.~J. and {Lu}, R. and {Lv}, H.~K. and {Ma}, B.~Q. and {Ma}, L.~L. and {Ma}, X.~H. and {Mao}, J.~R. and {Masood}, A. and {Mitthumsiri}, W. and {Montaruli}, T. and {Nan}, Y.~C. and {Pang}, B.~Y. and {Pattarakijwanich}, P. and {Pei}, Z.~Y. and {Qi}, M.~Y. and {Ruffolo}, D. and {Rulev}, V. and {S{\'a}iz}, A. and {Shao}, L. and {Shchegolev}, O. and {Sheng}, X.~D. and {Shi}, J.~R. and {Song}, H.~C. and {Stenkin}, Yu. V. and {Stepanov}, V. and {Sun}, Q.~N. and {Sun}, X.~N. and {Sun}, Z.~B. and {Tam}, P.~H.~T. and {Tang}, Z.~B. and {Tian}, W.~W. and {Wang}, B.~D. and {Wang}, C. and {Wang}, H. and {Wang}, H.~G. and {Wang}, J.~C. and {Wang}, J.~S. and {Wang}, L.~P. and {Wang}, L.~Y. and {Wang}, R.~N. and {Wang}, W. and {Wang}, W. and {Wang}, X.~G. and {Wang}, X.~J. and {Wang}, X.~Y. and {Wang}, Y.~D. and {Wang}, Y.~J. and {Wang}, Y.~P. and {Wang}, Z. and {Wang}, Z. and {Wang}, Z.~H. and {Wang}, Z.~X. and {Wei}, D.~M. and {Wei}, J.~J. and {Wei}, Y.~J. and {Wen}, T. and {Wu}, C.~Y. and {Wu}, H.~R. and {Wu}, S. and {Wu}, W.~X. and {Wu}, X.~F. and {Xi}, S.~Q. and {Xia}, J. and {Xia}, J.~J. and {Xiang}, G.~M. and {Xiao}, G. and {Xiao}, H.~B. and {Xin}, G.~G. and {Xin}, Y.~L. and {Xing}, Y. and {Xu}, D.~L. and {Xu}, R.~X. and {Xue}, L. and {Yan}, D.~H.},
        title = "{Observation of the Crab Nebula with LHAASO-KM2A - a performance study}",
      journal = {Chinese Physics C},
         year = 2021,
        month = feb,
       volume = {45},
       number = {2},
          eid = {025002},
        pages = {025002},
          doi = {10.1088/1674-1137/abd01b},
archivePrefix = {arXiv},
       eprint = {2010.06205},
 primaryClass = {astro-ph.HE},
       adsurl = {https://ui.adsabs.harvard.edu/abs/2021ChPhC..45b5002A}
}

@ARTICLE{HESS2018,
       author = {{H.~E.~S.~S. Collaboration} and {Abdalla}, H. and {Abramowski}, A. and {Aharonian}, F. and {Ait Benkhali}, F. and {Akhperjanian}, A.~G. and {Andersson}, T. and {Ang{\"u}ner}, E.~O. and {Arrieta}, M. and {Aubert}, P. and {Backes}, M. and {Balzer}, A. and {Barnard}, M. and {Becherini}, Y. and {Becker Tjus}, J. and {Berge}, D. and {Bernhard}, S. and {Bernl{\"o}hr}, K. and {Blackwell}, R. and {B{\"o}ttcher}, M. and {Boisson}, C. and {Bolmont}, J. and {Bordas}, P. and {Bregeon}, J. and {Brun}, F. and {Brun}, P. and {Bryan}, M. and {Bulik}, T. and {Capasso}, M. and {Carr}, J. and {Casanova}, S. and {Cerruti}, M. and {Chakraborty}, N. and {Chalme-Calvet}, R. and {Chaves}, R.~C.~G. and {Chen}, A. and {Chevalier}, J. and {Chr{\'e}tien}, M. and {Colafrancesco}, S. and {Cologna}, G. and {Condon}, B. and {Conrad}, J. and {Cui}, Y. and {Davids}, I.~D. and {Decock}, J. and {Degrange}, B. and {Deil}, C. and {Devin}, J. and {deWilt}, P. and {Dirson}, L. and {Djannati-Ata{\"\i}}, A. and {Domainko}, W. and {Donath}, A. and {Drury}, L.~O. 'C. and {Dubus}, G. and {Dutson}, K. and {Dyks}, J. and {Edwards}, T. and {Egberts}, K. and {Eger}, P. and {Ernenwein}, J.-P. and {Eschbach}, S. and {Farnier}, C. and {Fegan}, S. and {Fernandes}, M.~V. and {Fiasson}, A. and {Fontaine}, G. and {F{\"o}rster}, A. and {Fukuyama}, T. and {Funk}, S. and {F{\"u}{\ss}ling}, M. and {Gabici}, S. and {Gajdus}, M. and {Gallant}, Y.~A. and {Garrigoux}, T. and {Giavitto}, G. and {Giebels}, B. and {Glicenstein}, J.~F. and {Gottschall}, D. and {Goyal}, A. and {Grondin}, M.-H. and {Hadasch}, D. and {Hahn}, J. and {Haupt}, M. and {Hawkes}, J. and {Heinzelmann}, G. and {Henri}, G. and {Hermann}, G. and {Hervet}, O. and {Hinton}, J.~A. and {Hofmann}, W. and {Hoischen}, C. and {Holler}, M. and {Horns}, D. and {Ivascenko}, A. and {Jacholkowska}, A. and {Jamrozy}, M. and {Janiak}, M. and {Jankowsky}, D. and {Jankowsky}, F. and {Jingo}, M. and {Jogler}, T. and {Jouvin}, L. and {Jung-Richardt}, I. and {Kastendieck}, M.~A. and {Katarzy{\'n}ski}, K. and {Katz}, U. and {Kerszberg}, D. and {Kh{\'e}lifi}, B. and {Kieffer}, M. and {King}, J. and {Klepser}, S. and {Klochkov}, D. and {Klu{\'z}niak}, W. and {Kolitzus}, D. and {Komin}, Nu. and {Kosack}, K. and {Krakau}, S. and {Kraus}, M. and {Krayzel}, F. and {Kr{\"u}ger}, P.~P. and {Laffon}, H. and {Lamanna}, G. and {Lau}, J. and {Lees}, J.-P. and {Lefaucheur}, J. and {Lefranc}, V. and {Lemi{\`e}re}, A. and {Lemoine-Goumard}, M. and {Lenain}, J.-P. and {Leser}, E. and {Lohse}, T. and {Lorentz}, M. and {Liu}, R. and {L{\'o}pez-Coto}, R. and {Lypova}, I. and {Marandon}, V. and {Marcowith}, A. and {Mariaud}, C. and {Marx}, R. and {Maurin}, G. and {Maxted}, N. and {Mayer}, M. and {Meintjes}, P.~J. and {Meyer}, M. and {Mitchell}, A.~M.~W. and {Moderski}, R. and {Mohamed}, M. and {Mohrmann}, L. and {Mor{\r{a}}}, K. and {Moulin}, E. and {Murach}, T. and {de Naurois}, M. and {Niederwanger}, F. and {Niemiec}, J. and {Oakes}, L. and {O'Brien}, P. and {Odaka}, H. and {{\"O}ttl}, S. and {Ohm}, S. and {Ostrowski}, M. and {Oya}, I. and {Padovani}, M. and {Panter}, M. and {Parsons}, R.~D. and {Pekeur}, N.~W. and {Pelletier}, G. and {Perennes}, C. and {Petrucci}, P.-O. and {Peyaud}, B. and {Piel}, Q. and {Pita}, S. and {Poon}, H. and {Prokhorov}, D. and {Prokoph}, H. and {P{\"u}hlhofer}, G. and {Punch}, M. and {Quirrenbach}, A. and {Raab}, S. and {Reimer}, A. and {Reimer}, O. and {Renaud}, M. and {de los Reyes}, R. and {Rieger}, F. and {Romoli}, C. and {Rosier-Lees}, S. and {Rowell}, G. and {Rudak}, B. and {Rulten}, C.~B. and {Sahakian}, V. and {Salek}, D. and {Sanchez}, D.~A. and {Santangelo}, A. and {Sasaki}, M. and {Schlickeiser}, R. and {Sch{\"u}ssler}, F. and {Schulz}, A. and {Schwanke}, U. and {Schwemmer}, S. and {Settimo}, M.},
        title = "{H.E.S.S. observations of RX J1713.7-3946 with improved angular and spectral resolution: Evidence for gamma-ray emission extending beyond the X-ray emitting shell}",
      journal = {\aap},
         year = 2018,
        month = apr,
       volume = {612},
          eid = {A6},
        pages = {A6},
          doi = {10.1051/0004-6361/201629790},
archivePrefix = {arXiv},
       eprint = {1609.08671},
 primaryClass = {astro-ph.HE},
       adsurl = {https://ui.adsabs.harvard.edu/abs/2018A&A...612A...6H}
}

@ARTICLE{Anguner2023,
       author = {{Ang{\"u}ner}, E.~O.},
        title = "{Exploring the high energy frontiers of the Milky Way with ground-based gamma-ray astronomy: PeVatrons and the quest for the origin of Galactic cosmic-rays}",
      journal = {arXiv e-prints},
         year = 2023,
        month = may,
          eid = {arXiv:2305.12729},
        pages = {arXiv:2305.12729},
          doi = {10.48550/arXiv.2305.12729},
archivePrefix = {arXiv},
       eprint = {2305.12729},
 primaryClass = {astro-ph.HE},
       adsurl = {https://ui.adsabs.harvard.edu/abs/2023arXiv230512729A}
}

@ARTICLE{HAWC2025,
       author = {{Albert}, A. and {Alfaro}, R. and {Alvarez}, C. and {Andr{\'e}s}, A. and {Anita-Rangel}, E. and {Araya}, M. and {Arteaga-Vel{\'a}zquez}, J.~C. and {Arunbabu}, K.~P. and {Avila Rojas}, D. and {Ayala Solares}, H.~A. and {Babu}, R. and {Bangale}, P. and {Belmont-Moreno}, E. and {BenZvi}, S. and {Bernal}, A. and {Brisbois}, C. and {Caballero-Mora}, K.~S. and {Cabrera Mart{\'\i}nez}, J.~I. and {Capistr{\'a}n}, T. and {Carrami{\~n}ana}, A. and {Carre{\'o}n}, F. and {Casanova}, S. and {Col{\'\i}n-Farias}, A.~L., P.: Colmenero-Cesar and {Cotti}, U. and {Cotzomi}, J. and {Couti{\~n}o de Le{\'o}n}, S. and {De la Fuente}, E. and {de Le{\'o}n}, C. and {DeLay}, R. and {Depaoli}, D. and {Desiati}, P. and {Di Lalla}, N. and {Diaz Hernandez}, R. and {Dingus}, B.~L. and {DuVernois}, M.~A. and {D{\'\i}az-V{\'e}lez}, J.~C. and {Ellsworth}, R.~W. and {Engel}, K. and {Ergin}, T. and {Espinoza}, C. and {Fan}, K.~L. and {Fang}, K. and {Fleischhack}, H. and {Fraija}, N. and {Fraija}, S. and {Garc{\'\i}a-Gonz{\'a}lez}, J.~A. and {Garfias}, F. and {Ghosh}, N. and {Goksu}, H. and {Gonzalez Mu{\~n}oz}, A. and {Gonz{\'a}lez}, M.~M. and {Gonz{\'a}lez}, J.~A. and {Goodman}, J.~A. and {Groetsch}, S. and {Gyeong}, J. and {Harding}, J.~P. and {Hernandez Cadena}, S. and {Herzog}, I. and {Huang}, D. and {Hueyotl-Zahuantitla}, F. and {H{\"u}ntemeyer}, P. and {Iriarte}, A. and {Kaufmann}, S. and {Kelley-Hoskins}, N. and {Kieda}, D. and {Lara}, A. and {Lauer}, R. and {Leavitt}, K. and {Lee}, W.~H. and {Lee}, J. and {Le{\'o}n Vargas}, H. and {Lewis}, T. and {Linnemann}, J.~T. and {Longinotti}, A.~L. and {Luis-Raya}, G. and {Malone}, K. and {Marinelli}, A. and {Marinelli}, S. and {Martinez}, O. and {Martinez-Castellanos}, I. and {Mart{\'\i}nez-Castro}, J. and {Mart{\'\i}nez-Huerta}, H. and {Matthews}, J.~A. and {McEnery}, J. and {Miranda-Romagnoli}, P. and {Mir{\'o}n-Enriquez}, P.~E. and {Montaruli}, T. and {Morales-Soto}, J.~A. and {Moreno}, E. and {Mostaf{\'a}}, M. and {Najafi}, M. and {Nayerhoda}, A. and {Nellen}, L. and {Nisa}, M.~U. and {Noriega-Papaqui}, R. and {Omodei}, N. and {Osorio-Archila}, M. and {Ponce}, E. and {P{\'e}rez Araujo}, Y. and {P{\'e}rez-P{\'e}rez}, E.~G. and {Pretz}, J. and {Rho}, C.~D. and {Rodriguez Parra}, A. and {Rosa-Gonz{\'a}lez}, D. and {Roth}, M. and {Ruiz-Velasco}, E. and {Ryan}, J. and {Salazar}, H. and {Salazar-Gallegos}, D. and {Salesa Greus}, F. and {Sandoval}, A. and {Schneider}, M. and {Schwefer}, G. and {Serna-Franco}, J. and {Smith}, A.~J. and {Son}, Y. and {Springer}, R.~W. and {Taboada}, I. and {Tibolla}, O. and {Tollefson}, K. and {Torres}, I. and {Torres-Escobedo}, R. and {Turner}, R. and {Ure{\~n}a-Mena}, F. and {Varela}, E. and {Vargas-Maga{\~n}a}, M. and {V{\'a}zquez Estrada}, O.~W. and {Vianello}, G. and {Villase{\~n}or}, X., Wang and {Wang}, Z. and {Watson}, I.~J. and {Wisher}, I. and {Wood}, J. and {Wu}, H. and {Yapici}, T. and {Yu}, S. and {Yun-C{\'a}rcamo}, S. and {Zaborov}, D. and {Zhou}, H.},
        title = "{The first 10 years of the HAWC Gamma-Ray Observatory: science results}",
      journal = {\rmxaa},
         year = 2025,
        month = nov,
       volume = {61},
        pages = {261-299},
          doi = {10.22201/ia.01851101p.2025.61.03.13},
archivePrefix = {arXiv},
       eprint = {2512.03938},
 primaryClass = {astro-ph.HE},
       adsurl = {https://ui.adsabs.harvard.edu/abs/2025RMxAA..61c.261L}
}

@ARTICLE{Bahmanabadi2018,
       author = {{Bahmanabadi}, Mahmud and {Moghaddam}, Saba Mortazavi},
        title = "{Simulation of the time structure of Extensive Air Showers with CORSIKA initiated by various primary particles at Alborz-I observatory level}",
      journal = {\na},
         year = 2018,
        month = may,
       volume = {61},
        pages = {5-13},
          doi = {10.1016/j.newast.2017.11.003},
       adsurl = {https://ui.adsabs.harvard.edu/abs/2018NewA...61....5B}
}

@ARTICLE{Bassi1953,
       author = {{Bassi}, P. and {Clark}, G. and {Rossi}, B.},
        title = "{Distribution of Arrival Times of Air Shower Particles}",
      journal = {Physical Review},
         year = 1953,
        month = oct,
       volume = {92},
       number = {2},
        pages = {441-451},
          doi = {10.1103/PhysRev.92.441},
       adsurl = {https://ui.adsabs.harvard.edu/abs/1953PhRv...92..441B}
}

@ARTICLE{Bonifazi2009,
       author = {{Bonifazi}, C. and {Pierre Auger Collaboration}},
        title = "{The angular resolution of the Pierre Auger Observatory}",
      journal = {Nuclear Physics B Proceedings Supplements},
         year = 2009,
        month = may,
       volume = {190},
        pages = {20-25},
          doi = {10.1016/j.nuclphysbps.2009.03.063},
archivePrefix = {arXiv},
       eprint = {0901.3138},
 primaryClass = {astro-ph.HE},
       adsurl = {https://ui.adsabs.harvard.edu/abs/2009NuPhS.190...20B}
}

@ARTICLE{Cao2023,
       author = {{Cao}, Zhen and {Chen}, Songzhan and {Liu}, Ruoyu and {Yang}, Ruizhi},
        title = "{Ultra-High-Energy Gamma-Ray Astronomy}",
      journal = {Annual Review of Nuclear and Particle Science},
         year = 2023,
        month = sep,
       volume = {73},
        pages = {341-363},
          doi = {10.1146/annurev-nucl-112822-025357},
archivePrefix = {arXiv},
       eprint = {2310.01744},
 primaryClass = {astro-ph.HE},
       adsurl = {https://ui.adsabs.harvard.edu/abs/2023ARNPS..73..341C}
}

@ARTICLE{Cao2021,
       author = {{Cao}, Zhen},
        title = "{EAS Arrays at High Altitudes Start the Era of UHE {\ensuremath{\gamma}}-ray Astronomy}",
      journal = {Universe},
         year = 2021,
        month = sep,
       volume = {7},
       number = {9},
          eid = {339},
        pages = {339},
          doi = {10.3390/universe7090339},
       adsurl = {https://ui.adsabs.harvard.edu/abs/2021Univ....7..339C}
}

@ARTICLE{CG2023,
       author = {{Cardillo}, Martina and {Giuliani}, Andrea},
        title = "{The LHAASO PeVatron bright sky: what we learned}",
      journal = {Applied Sciences},
         year = 2023,
        month = may,
       volume = {13},
       number = {11},
          eid = {6433},
        pages = {6433},
          doi = {10.3390/app13116433},
archivePrefix = {arXiv},
       eprint = {2305.10526},
 primaryClass = {astro-ph.HE},
       adsurl = {https://ui.adsabs.harvard.edu/abs/2023AppSc..13.6433C}
}

@ARTICLE{Luque2023,
       author = {{De La Torre Luque}, P. and {Gaggero}, D. and {Grasso}, D. and {Fornieri}, O. and {Egberts}, K. and {Steppa}, C. and {Evoli}, C.},
        title = "{Galactic diffuse gamma rays meet the PeV frontier}",
      journal = {\aap},
         year = 2023,
        month = apr,
       volume = {672},
          eid = {A58},
        pages = {A58},
          doi = {10.1051/0004-6361/202243714},
archivePrefix = {arXiv},
       eprint = {2203.15759},
 primaryClass = {astro-ph.HE},
       adsurl = {https://ui.adsabs.harvard.edu/abs/2023A&A...672A..58D}
}

@ARTICLE{Gabici2019,
       author = {{Gabici}, Stefano and {Evoli}, Carmelo and {Gaggero}, Daniele and {Lipari}, Paolo and {Mertsch}, Philipp and {Orlando}, Elena and {Strong}, Andrew and {Vittino}, Andrea},
        title = "{The origin of Galactic cosmic rays: Challenges to the standard paradigm}",
      journal = {International Journal of Modern Physics D},
         year = 2019,
        month = jan,
       volume = {28},
       number = {15},
          eid = {1930022-339},
        pages = {1930022-339},
          doi = {10.1142/S0218271819300222},
archivePrefix = {arXiv},
       eprint = {1903.11584},
 primaryClass = {astro-ph.HE},
       adsurl = {https://ui.adsabs.harvard.edu/abs/2019IJMPD..2830022G}
}

@BOOK{corsika,
       author = {{Heck}, D. and {Knapp}, J. and {Capdevielle}, J.~N. and {Schatz}, G. and {Thouw}, T.},
        title = "{CORSIKA: a Monte Carlo code to simulate extensive air showers.}",
         year = 1998,
       adsurl = {https://ui.adsabs.harvard.edu/abs/1998cmcc.book.....H}
}

@ARTICLE{Hofmann2020,
       author = {{Hofmann}, Werner},
        title = "{On angular resolution limits for air shower arrays}",
      journal = {Astroparticle Physics},
         year = 2020,
        month = dec,
       volume = {123},
          eid = {102479},
        pages = {102479},
          doi = {10.1016/j.astropartphys.2020.102479},
       adsurl = {https://ui.adsabs.harvard.edu/abs/2020APh...12302479H}
}

@INPROCEEDINGS{Huang2019,
       author = {{Huang}, D.},
        title = "{Studies of the time structure of extended air showers for direction reconstruction with the HAWC Outrigger array}",
    booktitle = {36th International Cosmic Ray Conference (ICRC2019)},
         year = 2019,
       series = {International Cosmic Ray Conference},
       volume = {36},
        month = jul,
          eid = {700},
        pages = {700},
          doi = {10.22323/1.358.0700},
archivePrefix = {arXiv},
       eprint = {1908.11456},
 primaryClass = {astro-ph.HE},
       adsurl = {https://ui.adsabs.harvard.edu/abs/2019ICRC...36..700H}
}

@INPROCEEDINGS{Inoue1985,
       author = {{Inoue}, N. and {Kawamoto}, M. and {Misaki}, Y. and {Maeda}, T. and {Takeuchi}, T. and {Toyoda}, Y.},
        title = "{Structure of Air Shower Disk Near the Core}",
    booktitle = {19th International Cosmic Ray Conference (ICRC19), Volume 7},
         year = 1985,
       editor = {{Jones}, F.~C.},
       series = {International Cosmic Ray Conference},
       volume = {7},
        month = aug,
        pages = {316},
       adsurl = {https://ui.adsabs.harvard.edu/abs/1985ICRC....7..316I}
}

@ARTICLE{Ivanov2021,
       author = {{Ivanov}, D. and {Kalashev}, O.~E. and {Kuznetsov}, M. Yu. and {Rubtsov}, G.~I. and {Sako}, T. and {Tsunesada}, Y. and {Zhezher}, Y.~V.},
        title = "{Using Deep Learning to Enhance Event Geometry Reconstruction for the Telescope Array Surface Detector}",
      journal = {arXiv e-prints},
         year = 2020,
        month = may,
          eid = {arXiv:2005.07117},
        pages = {arXiv:2005.07117},
          doi = {10.48550/arXiv.2005.07117},
archivePrefix = {arXiv},
       eprint = {2005.07117},
 primaryClass = {astro-ph.HE},
       adsurl = {https://ui.adsabs.harvard.edu/abs/2020arXiv200507117I}
}

@ARTICLE{minuit,
       author = {{James}, F. and {Roos}, M.},
        title = "{Minuit - a system for function minimization and analysis of the parameter errors and correlations}",
      journal = {Computer Physics Communications},
         year = 1975,
        month = dec,
       volume = {10},
       number = {6},
        pages = {343-367},
          doi = {10.1016/0010-4655(75)90039-9},
       adsurl = {https://ui.adsabs.harvard.edu/abs/1975CoPhC..10..343J}
}

@ARTICLE{Lafebre2009,
       author = {{Lafebre}, S. and {Engel}, R. and {Falcke}, H. and {H{\"o}randel}, J. and {Huege}, T. and {Kuijpers}, J. and {Ulrich}, R.},
        title = "{Universality of electron-positron distributions in extensive air showers}",
      journal = {Astroparticle Physics},
         year = 2009,
        month = apr,
       volume = {31},
       number = {3},
        pages = {243-254},
          doi = {10.1016/j.astropartphys.2009.02.002},
archivePrefix = {arXiv},
       eprint = {0902.0548},
 primaryClass = {astro-ph.HE},
       adsurl = {https://ui.adsabs.harvard.edu/abs/2009APh....31..243L}
}

@ARTICLE{LHAASO2024Bubble,
       author = {{Lhaaso Collaboration}},
        title = "{An ultrahigh-energy {\ensuremath{\gamma}} -ray bubble powered by a super PeVatron}",
      journal = {Science Bulletin},
         year = 2024,
        month = feb,
       volume = {69},
       number = {4},
        pages = {449-457},
          doi = {10.1016/j.scib.2023.12.040},
archivePrefix = {arXiv},
       eprint = {2310.10100},
 primaryClass = {astro-ph.HE},
       adsurl = {https://ui.adsabs.harvard.edu/abs/2024SciBu..69..449L}
}

@INPROCEEDINGS{Linsley1985,
       author = {{Linsley}, J.},
        title = "{Angular resolution of air shower array-telescopes.}",
    booktitle = {19th International Cosmic Ray Conference (ICRC19), Volume 3},
         year = 1985,
       editor = {{Jones}, F.~C.},
       series = {International Cosmic Ray Conference},
       volume = {3},
        month = aug,
        pages = {461-464},
       adsurl = {https://ui.adsabs.harvard.edu/abs/1985ICRC....3..461L}
}

@ARTICLE{Linsley1986,
       author = {{Linsley}, J.},
        title = "{Thickness of the particle swarm in cosmic-ray air showers}",
      journal = {Journal of Physics G Nuclear Physics},
         year = 1986,
        month = jan,
       volume = {12},
       number = {1},
        pages = {51-57},
          doi = {10.1088/0305-4616/12/1/013},
       adsurl = {https://ui.adsabs.harvard.edu/abs/1986JPhG...12...51L}
}

@ARTICLE{Teshima1986,
       author = {{Teshima}, T. and {Matsubara}, Y. and {Hara}, T. and {Hayashida}, N. and {Honda}, M. and {Ishikawa}, F. and {Kifune}, T. and {Mori}, M. and {Nagano}, M. and {Nishijima}, K. and {Ohoka}, H. and {Ohno}, Y. and {Tanahashi}, G.},
        title = "{Properties of {}10$^{9}$-{}10$^{10}$ GeV extensive air showers at core distances between 100 and 3000 m}",
      journal = {Journal of Physics G Nuclear Physics},
         year = 1986,
        month = oct,
       volume = {12},
       number = {10},
        pages = {1097-1113},
          doi = {10.1088/0305-4616/12/10/017},
       adsurl = {https://ui.adsabs.harvard.edu/abs/1986JPhG...12.1097T}
}

@INPROCEEDINGS{Zeng2025,
       author = {{Zeng}, J. and {Liu}, J. and {Zhang}, X. and {Zhang}, S. and {Wu}, S.},
        title = "{Angular resolution of LHAASO-KM2A with improved shower front profile and time structure}",
    booktitle = {39th International Cosmic Ray Conference},
         year = 2025,
        month = dec,
    publisher = {SISSA},
          eid = {449},
        pages = {449},
       adsurl = {https://ui.adsabs.harvard.edu/abs/2025icrc.confE.449Z}
}

@article{Cao2024,
    collaboration = {LHAASO Collaboration},
    author = "{LHAASO Collaboration}",
    title = "{The First LHAASO Catalog of Gamma-Ray Sources}",
    journal = {\apjs},
    year = 2024,
    volume = {271},
    number = {1},
    pages = {25},
    doi = {10.3847/1538-4365/acfd29},
    archivePrefix = {arXiv},
    eprint = {2305.17030},
    primaryClass = {astro-ph.HE},
    adsurl = {https://ui.adsabs.harvard.edu/abs/2024ApJS..271...25C},
}

@article{Kamata1958,
    author = {Kamata, Koichi and Nishimura, Jun},
    title = {The Lateral and the Angular Structure Functions of Electron Showers},
    journal = {Progress of Theoretical Physics Supplement},
    volume = {6},
    pages = {93--155},
    year = {1958},
    month = {February},
    doi = {10.1143/PTPS.6.93},
    adsurl = {https://ui.adsabs.harvard.edu/abs/1958PThPS...6...93K}
}

@article{Greisen1960,
    author = {Greisen, K.},
    title = {Cosmic Ray Showers},
    journal = {Annual Review of Nuclear Science},
    volume = {10},
    pages = {63--108},
    year = {1960},
    month = {December},
    doi = {10.1146/annurev.ns.10.120160.000431}
}

@article{Miyake1961,
    author = {Miyake, S. and Hinotani, K. and Ito, N. and Kino, S. and Sasaki, H. and Yoshii, H. and Sakuyama, H. and Kato, E.},
    title = {Lateral Density Distribution of Charged Particles in the Central Region of Extensive Air Shower},
    journal = {Journal of the Physical Society of Japan},
    volume = {16},
    number = {5},
    pages = {847--854},
    year = {1961},
    month = {May},
    doi = {10.1143/JPSJ.16.847}
}
	\bibliographystyle{JHEP}
	
\end{document}